\documentclass[%
reprint,
amsmath,amssymb,
aps,
]{revtex4-1}

\usepackage{array}
\newcolumntype{L}[1]{>{\raggedright\let\newline\\\arraybackslash\hspace{0pt}}m{#1}}
\newcolumntype{C}[1]{>{\centering\let\newline\\\arraybackslash\hspace{0pt}}m{#1}}
\newcolumntype{R}[1]{>{\raggedleft\let\newline\\\arraybackslash\hspace{0pt}}m{#1}}

\usepackage{color}

\usepackage{tabularx}
\usepackage{hyperref}
\usepackage{graphicx}
\usepackage{dcolumn}
\usepackage{bm}
\usepackage{amsmath}

\newcommand{\beginsupplement}{%
\setcounter{table}{0}
\renewcommand{\thetable}{S\Roman{table}}%
\setcounter{figure}{0}
\renewcommand{\thefigure}{S\arabic{figure}}%
}

\begin{document}

\preprint{APS/123-QED}

\title{Inter-regional ECoG correlations predicted\\ by communication dynamics, geometry, and correlated gene expression}

\author{Richard F. Betzel$^1$}
\author{John D. Medaglia$^2$}
\author{Ari E. Kahn$^{1,3,4}$}
\author{Jonathan Soffer$^1$}
\author{Daniel R. Schonhaut$^3$}
\author{Danielle S. Bassett$^{1,5}$}
\email{dsb @ seas.upenn.edu}
\affiliation{
	$^1$Department of Bioengineering, University of Pennsylvania, Philadelphia, PA, 19104 USA}
\affiliation{
	$^2$Department of Psychology, University of Pennsylvania, Philadelphia, PA, 19104 USA}
\affiliation{
	$^3$Department of Neuroscience, University of Pennsylvania, Philadelphia, PA, 19104 USA}
\affiliation{
	$^4$Human Research and Engineering Directorate, U.S. Army Research Laboratory, Aberdeen, MD 21001, USA}
\affiliation{
	$^5$Department of Electrical and Systems Engineering, University of Pennsylvania, Philadelphia, PA, 19104 USA}

\date{\today}

\begin{abstract}
Electrocorticography (ECoG) provides direct measurements of synchronized postsynaptic potentials at the exposed cortical surface. Patterns of signal covariance across ECoG sensors have been associated with diverse cognitive functions and remain a critical marker of seizure onset, progression, and termination. Yet, a systems level understanding of these patterns (or networks) has remained elusive, in part due to variable electrode placement and sparse cortical coverage. Here, we address these challenges by constructing inter-regional ECoG networks from multi-subject recordings, demonstrate similarities between these networks and those constructed from blood-oxygen-level-dependent signal in functional magnetic resonance imaging, and predict network topology from anatomical connectivity, interregional distance, and correlated gene expression patterns. Our models accurately predict out-of-sample ECoG networks and perform well even when fit to data from individual subjects, suggesting shared organizing principles across persons. In addition, we identify a set of genes whose brain-wide co-expression is highly correlated with ECoG network organization. Using gene ontology analysis, we show that these same genes are enriched for membrane and ion channel maintenance and function, suggesting a molecular underpinning of ECoG connectivity. Our findings provide fundamental understanding of the factors that influence interregional ECoG networks, and open the possibility for predictive modeling of surgical outcomes in disease.
\end{abstract}

\maketitle


\section*{Introduction}

Temporally correlated fluctuations in the spontaneous activity of cells, neuronal populations, and brain areas are hallmarks of neural systems and reflect their intrinsic functional organization \cite{friston1994functional, buzsaki2004neuronal}. These correlation patterns can be modeled as networks of functionally connected neural elements and analyzed using tools from network science \cite{bassett2017network}. Past studies have documented changes in functional network organization with cognitive state \cite{shirer2012decoding, gonzalez2015tracking}, in disease \cite{buckner2009cortical, fornito2015connectomics}, across individuals \cite{finn2015functional, gordon2016individual}, and over the human lifespan \cite{betzel2014changes, cao2014topological}, highlighting its utility for classification and diagnostic purposes.

Despite this, the neurobiological factors and architectural principles that shape baseline (normative) functional network organization remain elusive. Attempts to identify such principles empirically have been limited, largely, to networks reconstructed from the fMRI BOLD signal. While BOLD is a powerful, non-invasive technique for imaging the human brain \cite{logothetis2001neurophysiological}, it represents an indirect measure of neural activity and is further limited due to a slow, spatially non-uniform hemodynamic signal \cite{aguirre1998variability} that has proven especially susceptible to motion and vascular artifacts \cite{power2012spurious, liu2013neurovascular}. Moreover, the increased reliance on BOLD to study functional network organization could be seen as scientifically risky, potentially resulting in undo emphasis placed on organizational principles specific to BOLD while overlooking more general, cross-modal principles.

An appealing alternative to BOLD is electrocorticography (ECoG), which records electrical activity directly from the cortical surface in the form of synchronized postsynaptic potentials \cite{dringenberg1998involvement}. An invasive technique typically reserved for clinical settings, ECoG offers the same temporal resolution as scalp EEG (approximately 1 kHz) but with improved spatial specificity (signals are unimpeded by the skull). Despite its utility in marking abnormal neural dynamics and its link to cognitive phenomena \cite{gunduz2012decoding, voytek2015oscillatory, jacobs2016direct, branco2017decoding}, relatively few studies have investigated the network organization of ECoG beyond clinical contexts \cite{wilke2011graph, burns2014network, keller2014corticocortical, khambhati2016virtual, proix2017individual}.

Extending network analyses of ECoG \cite{ortega2008complex, kramer2010coalescence, chu2012emergence, he2008electrophysiological, hacker2017frequency} requires overcoming two critical limitations: (i) inconsistent electrode placement across subjects impedes population-level comparisons \cite{zumsteg2000presurgical}, and (ii) electrode coverage is typically restricted to select brain areas, precluding the possibility of whole-brain recordings. These limitations hamper the utility of ECoG in explaning both general principles of and individual differences in functional network organization. Developing novel tools and analytic methods to address these challenges would set the stage for fundamental studies identifying the factors that shape the correlation structure of the ECoG signal and aid in the prediction of clinical outcomes \cite{solomonwidespread}.

Here, we tackle these issues and propose a novel framework for constructing whole-brain, parcellation-based, and band-limited functional connectivity networks (ECoG FC) through the consolidation of multi-subject recordings. We show that ECoG FC shares a topological correspondence with networks reconstructed from fMRI BOLD, including correlated connection weights, distance and frequency dependence, as well as similar modular structure. Next, we identify factors that shape ECoG connectivity and use a multilinear model to predict ECoG connection weights on the basis of anatomical connectivity, inter-regional distance, and correlated gene expression patterns. We show that the most parsimonious models incorporate multiple factors to generate accurate predictions. In addition, we show that the performance of these models can be improved by computing gene expression correlation matrices using restricted subsets of genes. Importantly, these subsets are enriched for maintenance and regulation of ion channels and membrane potentials, suggesting genetic underpinnings of ECoG connectivity. Finally, we fit models to single-subject ECoG networks and show that the best-fitting models exhibit both a high degree of specificity (they generate the best predictions for the subject they were fit to) and a high degree of generalizability (they generate good out-of-sample predictions). Our results delineate factors that underpin the brain's intrinsic functional organization and in future work could be used to inform pre-surgical planning, e.g., by predicting the effect of cortical resection.

\section*{Results}

\subsection*{Whole-brain ECoG functional connectivity network}

\begin{figure*}[t]
	\begin{center}
		\centerline{\includegraphics[width=1\textwidth]{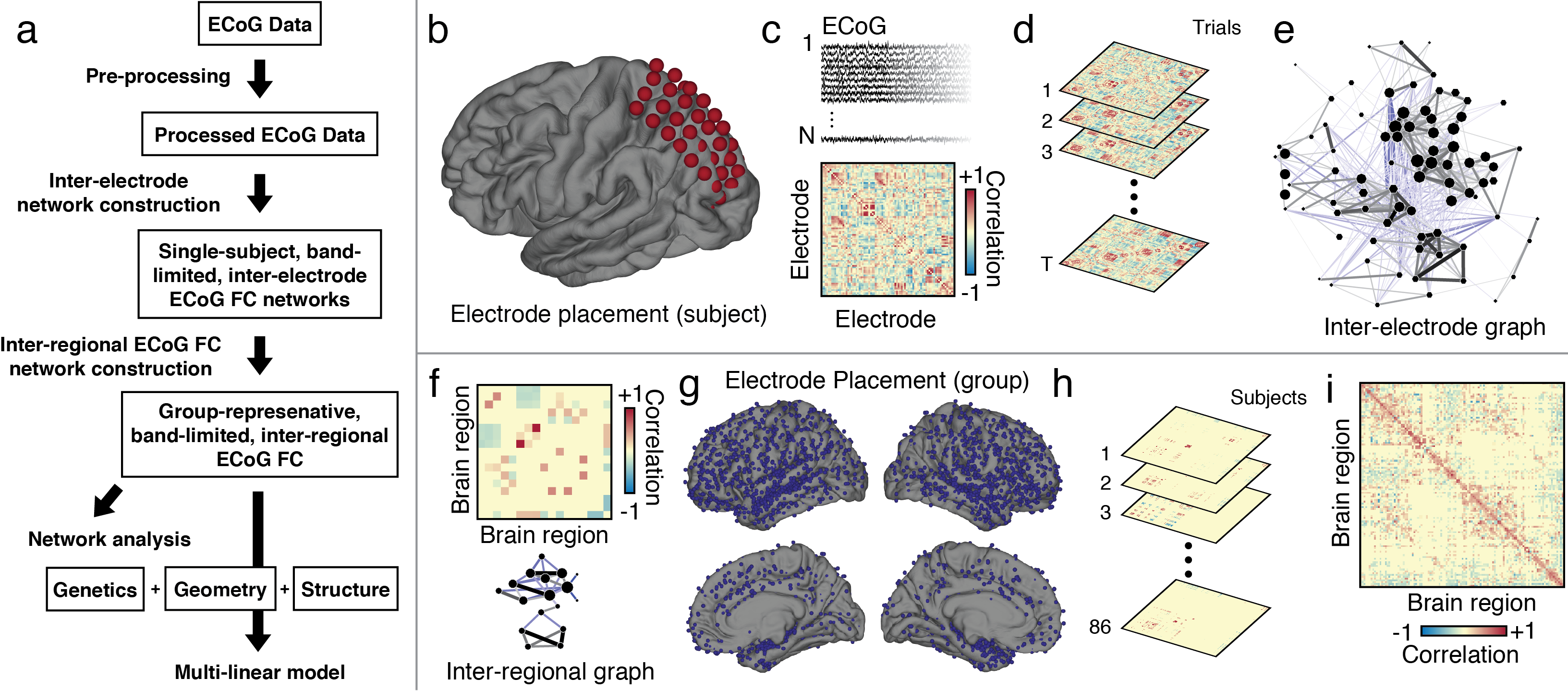}}
		\caption{\textbf{Processing pipeline for group-level ECoG functional connectivity (FC) matrices.} (\emph{a}) Schematic showing overall processing and analysis structure. (\emph{b}) Raw ECoG data were recorded from electrodes. (\emph{c}) The recordings were pre-processed and, for each trial and frequency band, we constructed a representative correlation matrix. (\emph{d}) We extracted connections that were consistently strong across all trials. (\emph{e}) The result of this procedure is a set of single-subject, band-limited, inter-electrode ECoG FC networks. (\emph{f}) We mapped electrode locations to vertices on the brain's surface and subsequently to brain regions. This procedure resulted in an inter-regional ECoG FC representation of each subject's inter-electrode network. (\emph{g}) By aggregating electrodes across the entire cohort, (\emph{h}) we were able to combine inter-regional FC networks to generate an estimate of whole-brain, inter-regional ECoG FC. (\emph{i}) From this aggregation procedure we calculated each connection's average weight across those observations.} \label{pipeline}
	\end{center}
\end{figure*}

We analyzed ECoG from 86 subjects recorded during resting periods between trials of a free recall task (Fig.~\ref{pipeline}). The process of estimating whole-brain functional connectivity (FC) from these recordings involved several steps. First, ECoG were pre-processed and filtered into seven frequency bands (1-4 Hz, 4-8 Hz, 8-13 Hz, 13-25 Hz, 25-45 Hz, 85-115 Hz, and 140-165 Hz). From the filtered time series we calculated for each subject and trial the full matrix of inter-electrode correlations, which was then transformed into an inter-regional correlation matrix by mapping electrodes to $N = 114$ brain regions based on their locations in MNI standard space \cite{cammoun2012mapping}. Finally, inter-regional matrices were averaged over trials and aggregated across subjects. This resulted in seven band-limited and group-representative correlation matrices, $\mathbf{A}^{\text{ECoG}} \in \mathbb{R}^{N \times N}$, whose element $A_{ij}^{\text{ECoG}}$ represented the average correlation (i.e., the functional connectivity; FC) of electrodes located near brain region $i$ with those located near region $j$. We refer to these matrices as ECoG FC throughout this report (See \textbf{Materials and Methods} for more details of network construction).

\subsubsection*{ECoG and BOLD connection weights are correlated}

Whole-brain FC networks estimated from fluctuations in resting-state fMRI BOLD signal are thought to reflect the brain's intrinsic architecture and vary systematically with cognitive state, disease, and development. However, the neurobiological underpinning of BOLD FC is unclear, as the BOLD signal represents an indirect measurement of neural activity and is susceptible to artifacts. ECoG, on the other hand, is a more direct measure of the brain's electrical activity and can therefore be used to validate findings and constrain hypotheses related to BOLD FC. In this section, we compare the network organization of ECoG FC with that of BOLD FC (See \textbf{Materials and Methods} for more information on fMRI BOLD acquisition and network construction).

First, as a coarse assessment of similarity, we computed the Pearson correlation between ECoG and BOLD FC connection weights (Fig.~\ref{fcProperties}a). We observed statistically significant correlations across all frequency bands (Fig.~\ref{fcProperties}b; $p < 10^{-15}$; FDR-controlled at $q = 0.05$ to account for multiple comparisons). The strongest correlation was observed in the slowest frequency band (1-4 Hz; $r_S = 0.37$, $p < 10^{-15}$; Fig.~\ref{fcProperties}c,d), suggesting that slow, coherent fluctuations in electrical activity may contribute to observed patterns of BOLD FC.

\subsubsection*{Functional connections are band-specific and distance-dependent}

Next, we tested the hypothesis that long-distance coordination of brain areas is supported by the correlation of frequency-specific electrical activity \cite{kopell2000gamma}. First, we asssessed whether the magnitude of ECoG FC was related to connection length (Euclidean distance). We observed a statistically significant inverse relationship between these two variables ($p < 10^{-15}$), indicating that the electrical activity of nearby brain regions is more correlated than distant regions (Fig.~\ref{fcProperties}f,g). While FC magnitude descreased with distance, we nonetheless wished to test whether any frequency bands differentially exhibited more or less strong, long-distance connections. To test whether this was the case, we imposed weight and length thresholds on ECoG FC and counted the number of jointly supra-threshold connections. Because what constitutes a ``strong'' and ``long'' connection is arbitrary, we tested a range of thresholds, from 80 - 120 mm for ``long'' and 0.0 - 0.5 for ``strong'' (typical output is shown in Fig.~\ref{fcProperties}g). Because they represented extremes in terms of the frequency content, we compared the number of supra-threshold connections in the slowest (1-4 Hz) and fastest (140-165 Hz) bands. In terms of raw counts, the slowest band always exhibited stronger and longer connections than the fastest band. To test whether these differences could be accounted for by chance, we compared them against a randomized model and expressed counts as $z$-scores with respect to the mean and standard error of the null distribution (See \textbf{Materials and Methods: Network null model}). For all threshold combinations, we found that the observed difference exceeded that which was expected under the random model ($z \ge 2.5$; FDR-controlled at $q = 0.05$ to account for multiple comparisons; Fig.~\ref{fcProperties}h). These results indicate that the slow frequency bands contain disproportionately many strong and long connections, suggesting that long-distance coordination among brain regions is related to slow, correlated electrical activity.

\subsubsection*{ECoG modules overlap with functional systems}

\begin{figure*}[t]
	\begin{center}
		\centerline{\includegraphics[width=0.8\textwidth]{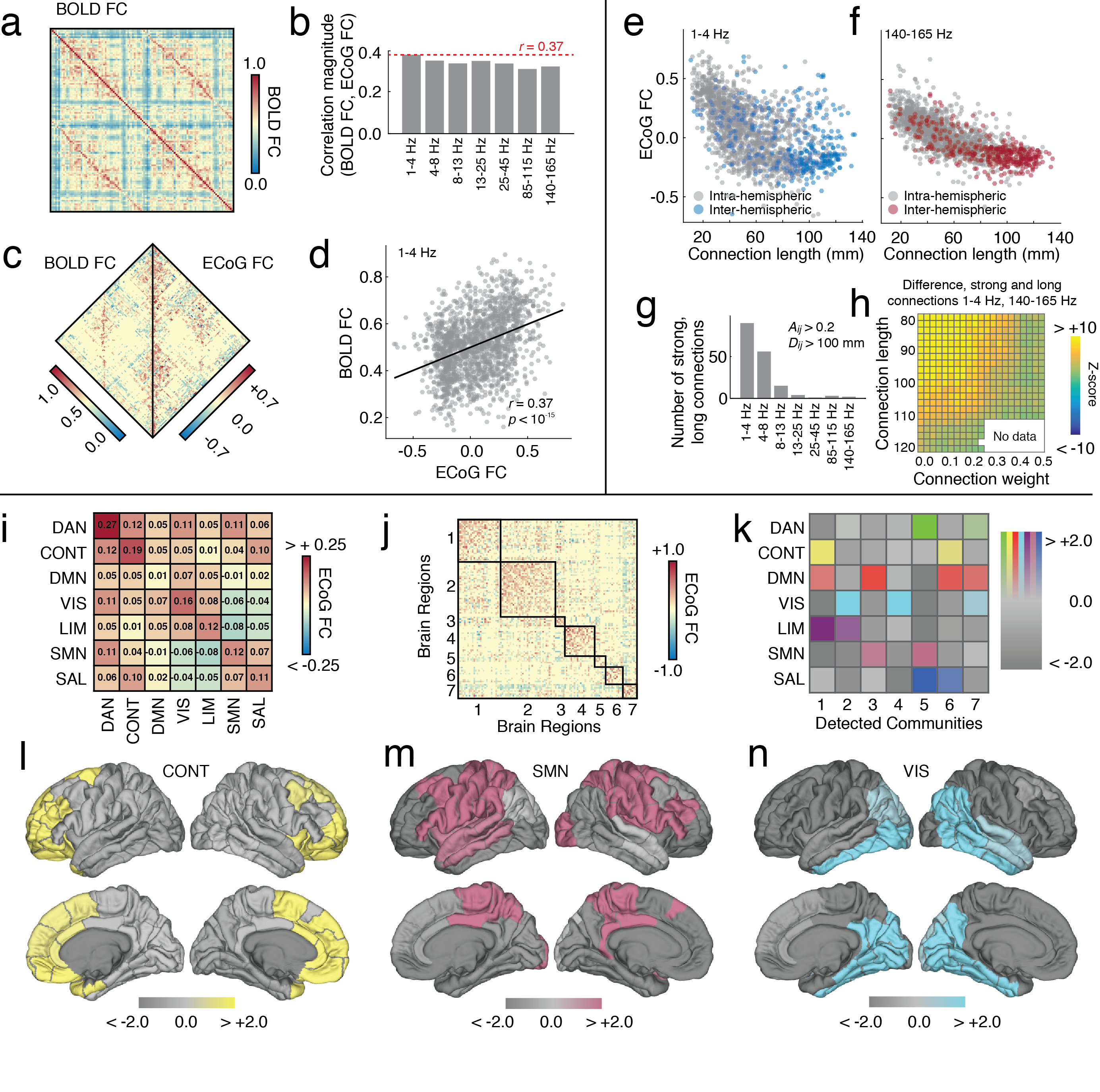}}
		\caption{\textbf{Relationship between group-level ECoG and BOLD functional connectivity.} (\emph{a}) Inter-regional BOLD correlation matrix. (\emph{b}) Pearson correlation between ECoG and BOLD connectivity as a function of frequency bands. Side-by-side (\emph{c}) and scatterplot (for 1-4 Hz) (\emph{d}) comparisons of group-averaged ECoG and BOLD connection weights. Edge weight, on average, decreases as a function of distance. We show examples for the 1-4 Hz (\emph{e}) and 140-165 Hz (\emph{f}) frequency bands. On average, slower frequency bands exhibited a greater number of connections that were both strong and long. (\emph{g}) A comparison between the number of these connections for a particular combination of edge weight and distance thresholds. (\emph{h}) We systematically varied the threshold values and compared (via $z$-score) the observed difference in long distance connections between the slowest (1-4 Hz) and fastest frequency bands (140-165 Hz) against what would be expected under a null model. (\emph{i}) We aggregated 1-4 Hz ECoG connectivity weights by functional systems identified in \cite{yeo2011organization} and obtained the average connection weight within (or between) all pairs of systems. (\emph{j}) We also performed community detection on the ECoG connectivity matrix to uncover seven communities of densely-interconnected brain regions. (\emph{k}) These communities exhibit non-random overlap with known functional systems: dorsal attention (DAN), control (CONT), default mode (DMN), visual (VIS), limbic (LIM), somatomotor (SMN), and salience (SAL) networks. We show the extent to which communities reconstitute three exemplar systems: (\emph{l}) cognitive control, (\emph{m}) somatomotor, and (\emph{n}) visual networks.} \label{fcProperties}
	\end{center}
\end{figure*}

\begin{figure*}[t]
\begin{center}
	\centerline{\includegraphics[width=0.4\textwidth]{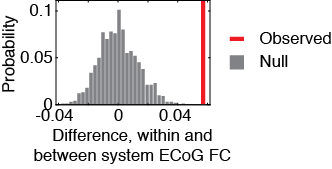}}
	\caption{\textbf{Mean within- and between-system ECoG FC magnitude.} We computed the difference between the average weights of connections within and between functional systems (red bar). We compared this value against a null distribution generated by permuting brain regions' system assignments at random (gray bars). We found that the observed difference far exceeded that which was expected by chance (1000 permutations, $p < 10^{-3}$).} \label{differenceBetweenWithinAndBetweenSystemECoGFC}
\end{center}
\end{figure*}

Lastly, we compared ECoG and BOLD FC in terms of their modular organization. Generally, modules -- which are internally dense and externally sparse subnetworks -- are thought to promote specialized information processing while conferring evolutionary adaptability to neural systems \cite{sporns2016modular}. In BOLD FC networks, modules overlap closely with the brain's task-activated functional systems \cite{power2011functional, yeo2011organization}. Here, we tested whether ECoG FC exhibits similar modular structure, focusing on the slowest frequency band (1-4 Hz) where the ECoG-BOLD FC correlation was strongest. We compared modules using two distinct strategies. First, we partitioned brain regions according to previously-defined functional system labels \cite{yeo2011organization} and calculated the mean within- \emph{versus} between-system connection weight (Fig.~\ref{fcProperties}i). Qualitatively, the system-by-system matrix shows a concentration of weight along the diagonal, suggesting that brain regions assigned to the same system tend to be more strongly connected to other regions in the same system than to regions in different systems. We quantitatively tested this observation and found that mean within-system weight of all systems combined exceeded that of between-system connections (permutation test, 10000 repetitions $p < 10^{-4}$, Fig.~\ref{differenceBetweenWithinAndBetweenSystemECoGFC}). This finding indicates that functional systems are topologically cohesive and further suggests that, at a system-level, correlation patterns among slowly fluctuating ECoG signals are relevant for the cognitive processes performed by systems commonly studied in resting state fMRI.

\begin{figure*}[t]
\begin{center}
	\centerline{\includegraphics[width=0.4\textwidth]{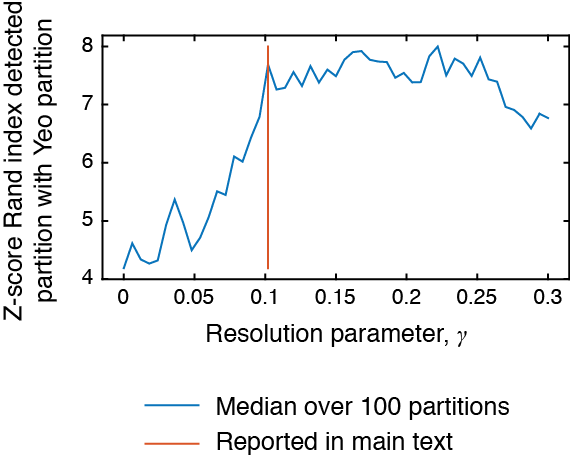}}
	\caption{\textbf{Similarity of detected modules with functional systems.} The blue line represents the mean similarity ($z$-score of the Rand coefficient) as a function of the resolution parameter, $\gamma$. We focused on partitions detected when similarity first plateaued (orange line).} \label{chooseGamma}
\end{center}
\end{figure*}

Rather than measure modularity using pre-defined system labels, we modified a well-known community detection algorithm to objectively partition ECoG FC into modules (see \textbf{Materials and Methods} for details) \cite{newman2004finding} and compared the overlap of module assignments with the \emph{a priori}-defined system labels (Fig.~\ref{fcProperties}j). The module detection process included a resolution parameter, $\gamma$, which scales the number and size of communities and whose value we set so that the detected modules were maximally similar to the pre-defined functional systems (Fig.~\ref{chooseGamma}) \cite{betzel2017modular}. The optimal partition consisted of seven modules, each of which exhibited nuanced and statistically significant overlap with functional systems (permutation test; $p < 0.01$; FDR-corrected) (Fig.~\ref{fcProperties}k). As an example, we show the overlap of detected modules with three systems (control, somatomotor, and visual) (Fig.~\ref{fcProperties}l-n). These findings offer an electrophysiological account of the organization of the cerebral cortex into functional systems, and lend further support for the hypothesis that the brain is well-described by spatially-distributed and topologically-cohesive modules.

\subsection*{Predicting whole-brain ECoG FC from geometry, structure, and genetics}

Despite the ease with which FC can be measured and accessed experimentally, it can be viewed epiphenomenally as the product of interacting structural, geometric, and genetic processes. Structural connections like synapses, axonal projections, and fiber bundles constrain communication patterns among neural elements and structure the propagation of activity across the brain and its correlation patterns \cite{honey2009predicting, adachi2011functional, goni2014resting}. Factors that influence anatomical connectivity also play important, albeit indirect, roles in shaping FC. The brain's intrinsic geometry and its drive to reduce metabolic and material connection costs result in wiring patterns that favor short, low-cost connections over longer, more costly connections \cite{bullmore2012economy, betzel2016generative}. Similarly, genetic factors regulate dendritic arborization \cite{bernard2010long} and myelin integrity \cite{chiang2009genetics, kochunov2011genetic}. Understanding how these and other factors shape functional network organization remains one of the overarching goals of network neuroscience \cite{bassett2017network}. While a number of studies have investigated how they relate to BOLD FC, virtually nothing is known about the relationship of these factors to networks estimated from ECoG.

\begin{figure*}[t]
	\begin{center}
		\centerline{\includegraphics[width=1\textwidth]{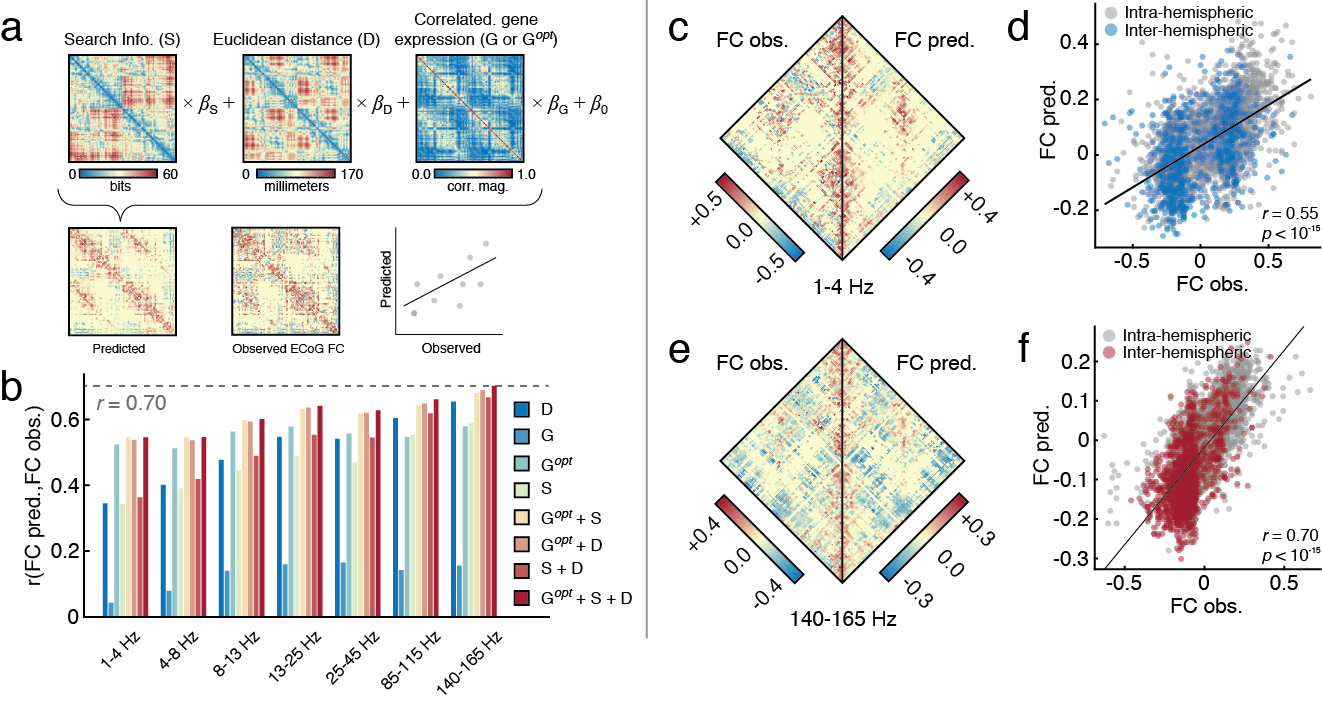}}
		\caption{\textbf{Predicting $A^\text{ECoG}$ with search information, Euclidean distance, and gene expression correlations.} Panel (\emph{a}) shows the general structure of the multi-linear model. Linear combinations of three predictors derived from brain structural connectivity (search information; $S$), spatial embedding (Euclidean distance; $D$), and genetics (gene expression correlation matrix or its optimized form; $G$ or $G^{\text{opt}}$) are used to generate predictions of ECoG connectivity. The regression coefficients, $\beta_S$, $\beta_D$, and $\beta_G$ are those that result in the maximum correlation of predicted and observed ECoG connectivity. We explore seven different combinations of predictors to identify the most parsimonious model. (\emph{b}) Correlation of predicted and observed ECoG connectivity for all seven models across all seven frequency bands. We show predicted and observed ECoG connectivity matrices side-by-side and as a scatterplot for the lowest frequency (1-4 Hz) (\emph{c,d}) and for the highest frequency bands (140-165 Hz) (\emph{e,f}). Gray points in the scatterplot represent intra-hemispheric connections while colored (blue; red) represent inter-hemispheric connections.} \label{multiLinearModel}
	\end{center}
\end{figure*}

To better understand how brain structure, geometry, and genetics influence ECoG FC, we investigated a set of nested multi-linear models (MLM) that generated predictions of ECoG FC connection weights, $\mathbf{\hat{A}}^{\text{ECoG}} = [\hat{A}_{ij}^{\text{ECoG}}]$. Predictions were made based on a linear combination of three predictors, each representing a different neurobiological mode capable of influencing ECoG FC: search information, $\mathbf{S} = [S_{ij}]$, which is computed from the matrix of reconstructed white-matter fiber pathways, measures the ``hiddenness" of the shortest anatomical path between regions $i$ and $j$ \cite{rosvall2005searchability, goni2014resting}; $\mathbf{D} = [D_{ij}]$; the Euclidean distance between regions $i$ and $j$; and $\mathbf{G} = [G_{ij}]$, the Pearson correlation between $i$ and $j$'s gene expression profiles (averaged across two donors) (see \textbf{Materials and Methods}). Model performance was defined as the Pearson correlation between the predicted ECoG FC and the observed ECoG FC. The full MLM including all three predictors is given by (Fig.~\ref{multiLinearModel}a):

\begin{equation}
\mathbf{A}^{\text{ECoG}} = \beta_0 + \beta_{S} \mathbf{S} + \beta_D \mathbf{D} + \beta_{G} \mathbf{G}.
\end{equation}

\noindent We tested all possible combinations of predictors, constituting seven models in total, and identified the optimal model for each frequency band based on the Akaike information criterion (AIC) \cite{akaike1998information}. For a given model, AIC was calculated as:

\begin{equation}
AIC = N_{samp} \log \bigg( \frac{RSS}{N_{samp}} \bigg) + 2 K\\,
\end{equation}

\noindent where $N_{samp}$, $RSS$, and $K$ were the total number of samples (pairs of brain regions for which ECoG FC information was available), residual sum of squared errors, and the total number of predictors (including the constant, $\beta_0$), respectively. The value of $N_{samp}$ was the same for all models (but varied with frequency band); models differed from one another only in terms of $RSS$ and $K$.

\subsubsection*{Single-factor models}

The simplest models we tested used single factors ($\mathbf{S}$, $\mathbf{D}$, or $\mathbf{G}$) to predict ECoG FC. Despite their simplicity, we found that in some cases they performed surprisingly well (Fig.~\ref{multiLinearModel}b; Table.~\ref{table1}). Across all frequency bands, search information and Euclidean distance performed the best. The correlation of predicted and observed ECoG FC based on either of these factors never fell below $r = 0.345$ and in the highest frequency bands reached a level of $r = 0.654$. Correlated gene expression, on the other hand, consistently performed worst, achieving a maximum correlation of $r = 0.156$.

These observations prompted us to pursue two additional experiments. First, because search information and Euclidean distance performed similarly and due to ongoing debate over the role that distance plays in shaping anatomical connectivity (used to estimate search information), we wanted to test whether search information generated statistically significant predictions of ECoG FC above and beyond that of Euclidean distance. As expected, we found that search information (derived from the network of white-matter connections) and distance were correlated with one another ($r = 0.74$, $p < 10^{-15}$). To assess search information's unique contribution to ECoG FC, we partialed out the effect of distance and used the residuals to predict ECoG FC. This analysis revealed that, while the overall magnitude of correlation is attenuated, the residuals nonetheless can account for some of the variance in ECoG FC (maximum $p \approx 1.3 \times 10^{-7}$; Fig.~\ref{regressDist}). In demonstrating a close correspondence between structural connectivity and distance, these results corroborate past studies that documented similar relationships. Also in line with past work, we show that search information (a measure based on structural connectivity) nonetheless makes a unique contribution in predicting ECoG FC beyond that of distance, alone.

\begin{figure*}[t]
	\begin{center}
		\centerline{\includegraphics[width=0.4\textwidth]{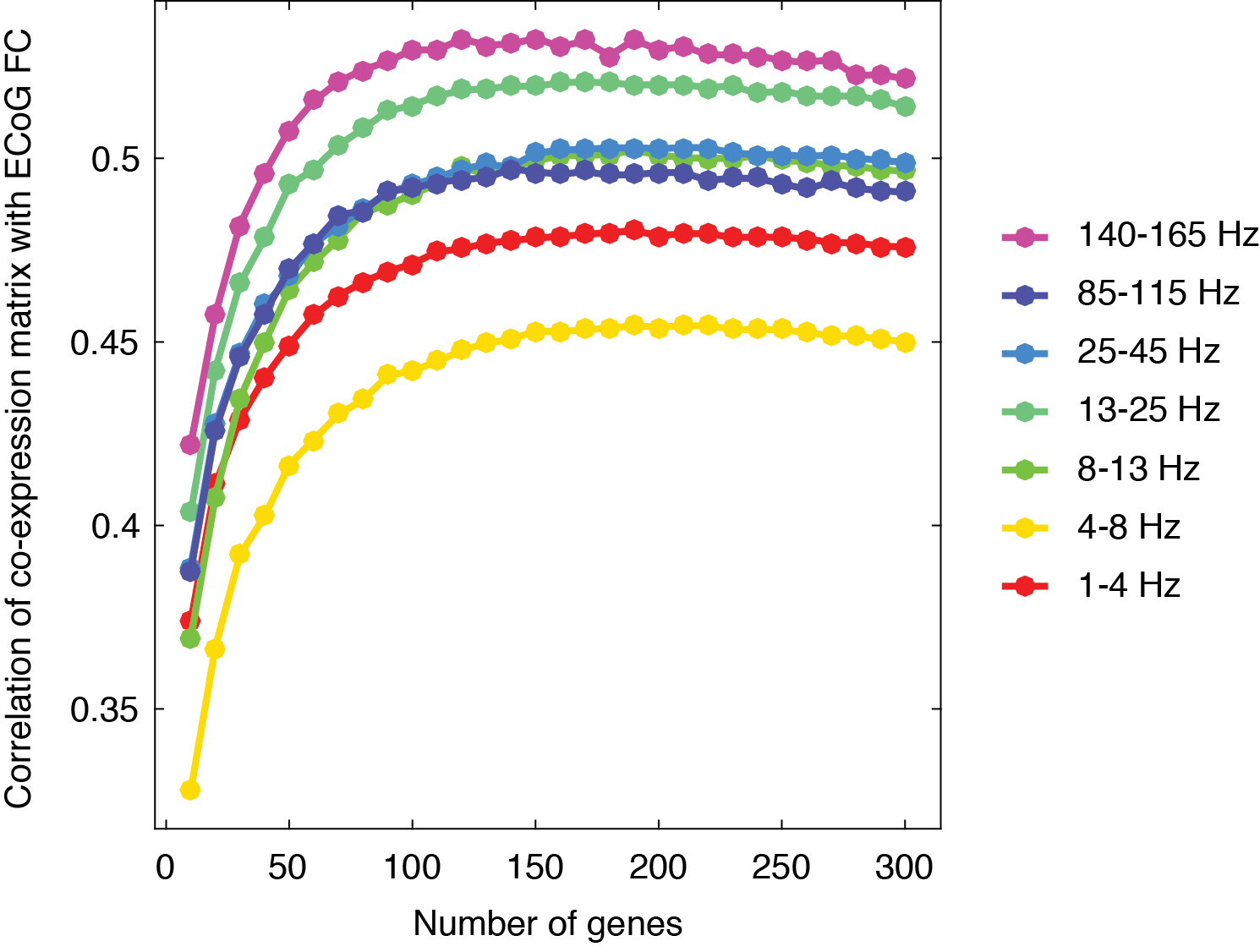}}
		\caption{\textbf{Correlation of optimized gene co-expression matrix with ECoG FC.} We used simulated annealing to identify sets of genes whose co-expression pattern was maximally correlated with ECoG FC. Here, we show the mean correlation (over 50 repeats of the optimization algorithm) as we vary the number of genes in the optimized list.} \label{chooseNumberOfGenes}
	\end{center}
\end{figure*}

Second, we wished to better understand why correlated gene expression performed so poorly in predicting ECoG FC. One hypothesis is that ECoG FC has little or no genetic basis. Past studies, however, have refuted this hypothesis \cite{richiardi2015correlated, krienen2016transcriptional}, consistently demonstrating a non-trivial relationship between genetics and FC, though mediated by small subsets of genes. This evidence prompted the alternative hypothesis that ECoG FC could be better predicted by shifting our focus away from the correlation patterns of $>10,000$ genes and narrowing our focus to the correlation patterns of small groups. Because the problem of identifying such groups is computationally intractable, we resorted to numerical methods to generate estimates. Briefly, we used a simulated annealing algorithm to optimize model performance while varying the size of the gene group (from 10 to 360 in increments of 10) and its membershp (See \textbf{Materials and Methods: Gene-ECoG Optimization} for more details; Fig.~\ref{chooseNumberOfGenes}). We repeated this procedure seperately for all seven frequency bands. We found that with groups of $181 \pm 23$ genes (mean and standard deviation across frequency bands), we could dramatically improve the model performance (Fig.~\ref{multiLinearModel}b; Table.\ref{table1}). Improvements were greatest in the slowest frequency band, with the performance of the genetics single-factor model increasing from $r = 0.043$ to $r = 0.523$. We refer to the correlation matrix of genes' expression profiles as $\mathbf{G}^{\text{opt}}$, indicating that the gene list was optimized to maximize its correspondence with ECoG FC. Note that in all subsequent analyses we use these optimized lists in place of the complete list of genes.

\subsubsection*{Multi-factor models}

In addition to the single-factor models, we also explored increasingly complex models, which included combinations of multiple factors. Seeking a balance between a model's explanatory power and its complexity, we used the Akaike information criterion (AIC) to identify the most parsimonious model for each frequency band. For the slowest frequency, the optimal model included two predictors (search information + optimized gene co-expression). For all other bands, the optimal model included all three predictors (search information + Euclidean distance + optimized gene co-expression), indicating that the brain's functional architecture, when estimated as ECoG FC, is shaped by a plurality of factors (Fig.~\ref{multiLinearModel}b; Table.~\ref{table1}). We show examples of predicted ECoG FC for the lowest (1-4 Hz) and highest (140-165 Hz) frequency bands (Fig.~\ref{multiLinearModel}c-f). It should also be noted that while all models tested here were fit using connections from across the entire brain, this framework can be easily extended to the level of individual brain systems, and it can be fit based on specific subsets of connections (See \textbf{Supplementary Materials: System-level multi-linear models}).

Collectively these findings build on past investigations into the singular roles played by structure, geometry, and genetics in shaping BOLD and ECoG FC \cite{chu2015eeg, goni2014resting, richiardi2015correlated}. While single-predictor models offered reasonable first approximations of ECoG FC, more complex models offered superior performance while maintaining parsimony. Interestingly, we found the search information and Euclidean distance had much greater explanatory power than the correlation pattern of all genes' expression levels. However, we also showed that the co-expression patterns of select subsets of genes were robustly related to ECoG FC, in agreement with past studies \cite{richiardi2015correlated, krienen2016transcriptional}.

The models we study here are exceedingly simple. Nonetheless, they represent the first attempt to identify the organizational principles and neurobiological factors that shape ECoG FC. These results are a natural extension of past studies that used similar techniques to model BOLD FC. However, while the BOLD signal is prone to motion \cite{power2012spurious}, respiratory \cite{birn2006separating}, and vascular\cite{liu2013neurovascular} artifacts, the ECoG signal is a relatively unimpeded measure of electrical activity, affording us greater confidence that the FC patters we analyze are indeed, of neuronal provenance.

\begin{table*}
	\begin{tabular}{|C{2cm}|C{1.5cm}|C{1.5cm}|C{1.5cm}|C{1.5cm}|C{1.5cm}|C{1.5cm}|C{1.5cm}|C{1.75cm}|}
		\hline
		\textbf{Model} & $D$ & $G$ & $G^{\text{opt.}}$ & $S$ & $G^{\text{opt.}},S$ & $G^{\text{opt.}},D$ & $S,D$ & $G^{\text{opt.}},D,S$ \\
		\hline
		1-4 Hz & $0.345$ & $0.043$ & $0.523$ & $0.343$ & $\mathbf{0.545}$ & $0.538$ & $0.363$ & $0.546$ \\
		\hline
		4-8 Hz & $0.401$ & $0.079$ & $0.512$ & $0.390$ & $0.545$ & $0.537$ & $0.418$ & $\mathbf{0.546}$ \\
		\hline
		8-13 Hz & $0.477$ & $0.140$ & $0.563$ & $0.446$ & $0.598$ & $0.593$ & $0.489$ & $\mathbf{0.601}$ \\
		\hline
		13-25 Hz & $0.547$ & $0.160$ & $0.578$ & $0.487$ & $0.631$ & $0.636$ & $0.553$ & $\mathbf{0.641}$ \\
		\hline
		25-45 Hz & $0.540$ & $0.165$ & $0.557$ & $0.470$ & $0.618$ & $0.620$ & $0.545$ & $\mathbf{0.627}$ \\
		\hline
		85-115 Hz & $0.604$ & $0.142$ & $0.546$ & $0.553$ & $0.641$ & $0.648$ & $0.618$ & $\mathbf{0.660}$ \\
		\hline
		140-165 Hz & $0.654$ & $0.156$ & $0.578$ & $0.589$ & $0.680$ & $0.689$ & $0.667$ & $\mathbf{0.702}$ \\
		\hline
	\end{tabular}
	\caption{\textbf{Model output.} Each column represents one of eight multi-linear models. The first row indicates which measures were used as predictors: $D$, $G$, $G^{\text{opt}}$, and $S$ represent Euclidean distance, gene co-expression, optimized gene co-expression, and search information. The next seven rows show the Pearson correlation magnitude of the predicted and empirical ECoG FC. The optimal models as determined by AIC are shown in boldface type.} \label{table1}
\end{table*}

\subsection*{Predicting single-subject ECoG FC}

\begin{figure*}[t]
	\begin{center}
		\centerline{\includegraphics[width=1\textwidth]{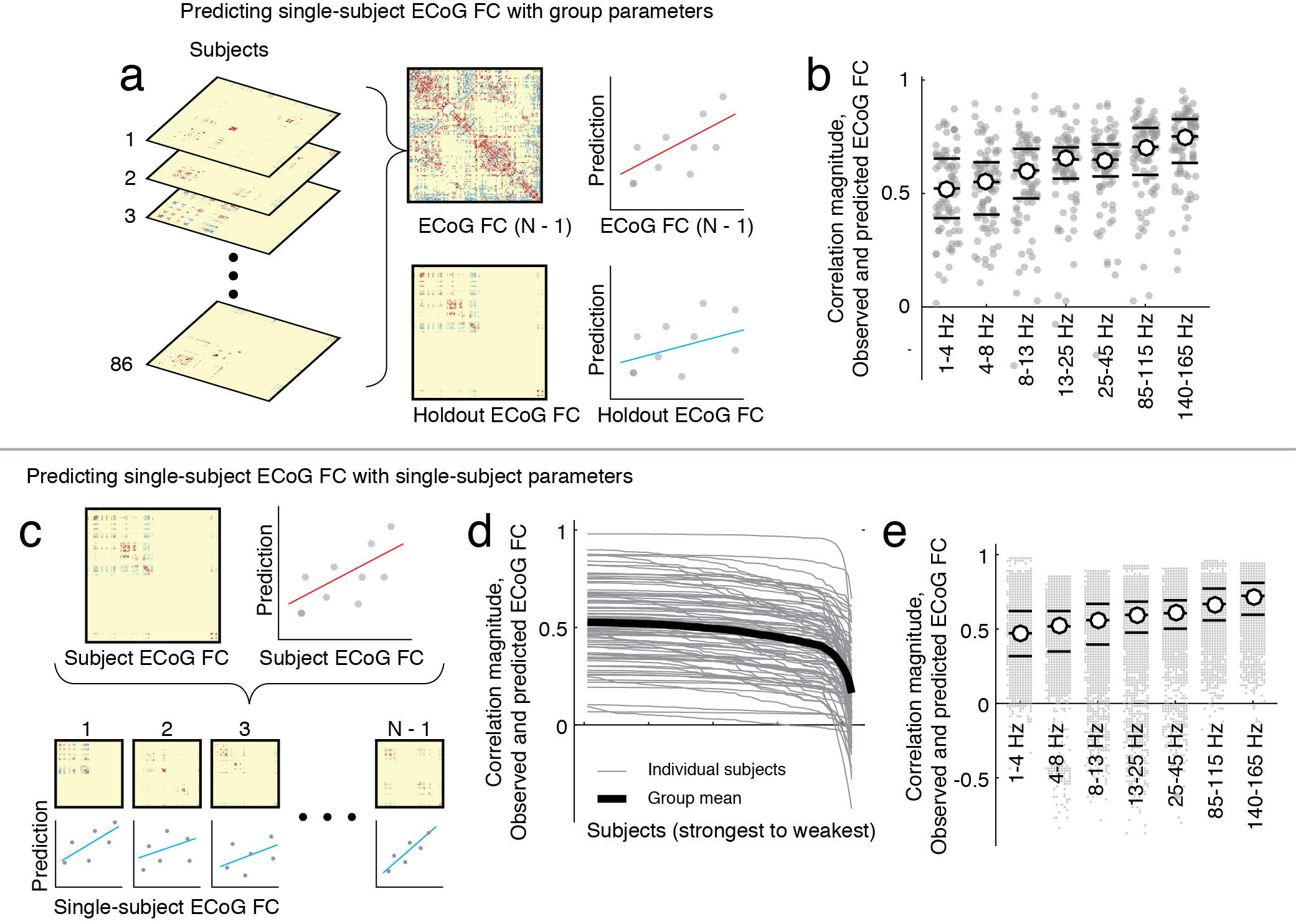}}
		\caption{\textbf{Predicting single-subject $A^{\text{ECoG}}$.} (\emph{a}) Schematic illustrating jackknife procedure -- models are fit using data from $N - 1$ subjects to predict the FC of the remaining subject. (\emph{b}) Correlation of predicted and observed single-subject ECoG FC as a function of frequency band. Each point represents a single subject. (\emph{c}) Schematic illustrating the single-subject model-fitting procedure -- models are fit to data from one subject and used to predict the FC of the remaining $N - 1$ subjects. (\emph{d}) Single-subject model performance. Each line represents the performance of a model fit to one of $N$ subjects. Model performance (Pearson correlation between observed and predicted ECoG FC) is ordered from best to worst. (\emph{e}) Correlation between predicted and observed single-subject ECoG FC as a function of frequency band.} \label{subjPredictions}
	\end{center}
\end{figure*}

To this point we have demonstrated that ECoG FC has properties similar to that of BOLD FC and that, with measures based on brain structure, geometry, and genetics, we can predict the magnitude of ECoG FC between brain regions. These analyses were carried out using group-representative data, which unfortunately makes it impossible to disentangle the contributions of individual subjects. Subject-level predictive models, on the other hand, have important clinical implications and open the possibility for predicting functional effects of neurosurgery or stimulation \cite{ezzyat2017direct}.

In this section, we extend the group-level predictive framework to the level of single-subject data. We show that the group-level models are generalizable and make good predictions of single-subject, out-of-sample ECoG FC. We also confront the more challenging task of fitting the model to incomplete single-subject data and, with the resulting models, predict the ECoG FC of other subjects. We find that the single-subject models exhibit stereotypical differences unique to each individual, but nonetheless remain highly generalizable and can predict the ECoG FC of other subjects. At the same time, demonstrating the generalizability of group-level models indicates that ECoG FC may be organized based on a shared set of principles.

First, we tested the group-level model's robustness using a jackknife procedure in which we estimated whole-brain ECoG FC matrices using data from $N - 1$ subjects. Next, we fit the full model using these data, and used the regression coefficients to predict the ECoG FC of the held out subject (Fig.~\ref{subjPredictions}a). We repeated this procedure, holding out each subject, and found that generally we could predict single-subject ECoG FC with a high degree of accuracy using the group-estimated regression coefficients. We observed that, for all frequency bands, the interquartile range of correlations between predicted and observed ECoG FC always excluded zero (Fig.~\ref{subjPredictions}b), demonstrating that the MLM approach has utility in predicting subject-level ECoG FC.

Using $N - 1$ subjects to fit model parameters and to estimate whole-brain ECoG FC is still relatively easy; the whole-brain inter-regional ECoG FC matrix contains thousands of observations used to fit the MLM, meaning that the optimal parameters are not especially biased by any single subject (which contributes to only a subset of the observations). A more challenging task is to fit the model using single-subject data, which offer far fewer observations of ECoG FC and are limited by the placement of electrode grids in terms of which inter-regional observations are available. Nonetheless, we tested whether models fit to individual subject's ECoG FC could be used to predict the ECoG FC of the remaining $N - 1$ subjects (Fig.~\ref{subjPredictions}c). If so, this would support the hypothesis that ECoG FC is organized according to similar wiring rules across different subjects.

We found that even with far fewer observations, we were still able to make good predictions of subjects' ECoG FC using parameters estimated from other subjects' ECoG FC. As expected, the parameter fits were subject-specific -- i.e., parameters best predicted ECoG FC of the subject from whose data they were estimated (Fig.~\ref{subjPredictions}d). Remarkably, however, the predictive capacity of these parameters did not immediately attenuate when they were applied to other subjects, with interquartiles ranges excluding zero as possible values (Fig.~\ref{subjPredictions}d). These findings suggest excellent generalizability and the possibility that similar organizational principles explain ECoG FC network architecture across subjects. Similar to previous sections, we observed that predictive capacity increased with frequency (Fig.~\ref{subjPredictions}e), suggesting that intersubject variability may be most pronounced in slower frequency bands.

\subsection*{Gene ontology analysis}

In the previous section, we found that when we calculated the correlation of gene expression profiles across the brain using $\approx 30000$ genes, the resulting matrix was weakly related to ECoG FC. Moreover, we found that by focusing on a small subset of genes we could dramatically improve this relationship. These findings are in line with past studies, in which the correlated expression levels of small subsets of genes ($\approx 10 - 100$) was found to be related to patterns of BOLD FC.

One risk associated with this approach is that, due to the number of genes, it might be trivial to find a small subset whose correlated expression profiles are similar to ECoG FC. In other words, optimizing an objective function could be effectively amplifying random fluctuations in a large dataset. One way to discount this possibility is to demonstrate that the genes that constitute the optimized list are not randomly selected and that, collectively, they comprise components of pathways that perform specific biological processes and cellular functions or encode for cellular components. To assess whether this was the case, we performed gene ontology (GO) analysis on the genes comprising the optimized list. We used the software GOrilla (\url{http://cbl-gorilla.cs.technion.ac.il}) to compare the optimized list of genes against the complete list of background genes \cite{eden2007discovering, eden2009gorilla}. We focus our analysis on the slowest frequency bands (1-4 Hz and 4-8 Hz) because the predictions of ECoG FC in these bands exhibited the greatest percent increase as a result of replacing the co-expression matrix calculated from the full set of genes with the corresponding matrix calculated from the optimized list.

In general, the GO analyses of both frequency bands resulted in similar findings, indicating that the optimized gene lists were enriched for biological functions related to the transport of ions across channels and cellular membranes. Near the top of both lists were ontology terms for ``sodium ion transport'', ``membrane depolarization during action potential'', ``monovalent inorganic cation transport'', ``regulation of transport'', ``sodium ion transmembrane transport'', and ``sodium ion transport'' (maximum $p$-value, $p =8.5 \times 10^{-4}$). Similarly, in terms of molecular function, both frequency bands were enriched for ``voltage-gated ion channel activity involved in regulation of postsynaptic membrane potential'' and ``voltage-gated sodium channel activity'' (maximum $p$-value, $p = 4.79 \times 10^{-4}$). Also, in terms of cellular components, the 4-8 Hz frequency band was enriched for terms related to membrane channels including ``cation channel complex'', ``voltage-gated sodium channel complex'', ``ion channel complex'', ``transmembrane transporter complex'', ``potassium channel complex'', ``transporter complex'', and ``sodium channel complex'' (maximum $p$-value, $p = 4.67 \times 10^{-4}$) (See \textbf{Supplementary Tables}.~\ref{freq1function}-.~\ref{freq2component} for a complete list of enriched terms).

In a previous section we demonstrated that the correspondence of ECoG FC and patterns of correlated gene expression could be strengthened by narrowing our focus onto select subsets of genes. Here, we offer additional support to futher strengthen this relationship, demonstrating that the optimized list of genes is enriched for terms associated with membrane channels and ion transport. These findings further suggest a molecular and genetic underpinning of ECoG FC.

\begin{table*}
		\begin{tabular}{|C{5cm}|C{2cm}|C{2cm}|C{2cm}|C{2cm}|C{2cm}|}
			\hline
			Frequency/Percentile & 5 & 25 & 50 & 75 & 95 \\
			\hline
			1-4 Hz & 0.212 & 0.391 & 0.522 & 0.654 & 0.795 \\
			\hline
			4-8 Hz & 0.208 & 0.407 & 0.550 & 0.638 & 0.794 \\
			\hline
			8-13 Hz & 0.158 & 0.478 & 0.602 & 0.696 & 0.800 \\
			\hline
			13-25 Hz & 0.263 & 0.565 & 0.650 & 0.703 & 0.853 \\
			\hline
			25-45 Hz & 0.269 & 0.575 & 0.649 & 0.716 & 0.818 \\
			\hline
			85-115 Hz & 0.330 & 0.582 & 0.705 & 0.789 & 0.860 \\
			\hline
			140-165 Hz & 0.378 & 0.635 & 0.751 & 0.827 & 0.887 \\
			\hline
		\end{tabular}
		\caption{\textbf{Results of jackknife procedure.} Each row represents one of seven frequency bands. The columns represent percentiles of correlation coefficient distributions. These distributions were obtained using a jackknife procedure that entailed using the multi-linear model fit built on data from $N - 1$ subjects to predict inter-regional ECoG FC of a held-out subject. The correlation coefficients measure the magnitude correlation of that subject's predicted and observed FC.} \label{table2}
\end{table*}
\subsection*{Robustness to methodological variation}

The results presented here depended upon a particular sequence of decisions concerning how to process, analyze, and synthesize several multi-modal brain imaging datasets. To ensure their robustness, we confirmed that our results hold under reasonable variation to this sequence. Specifically, we demonstrated the consistency of ECoG FC networks with respect to variation in the distance threshold used in the electrode-to-region mapping (Fig.~\ref{variationOfDistanceThreshold}) and using different measures of FC, namely phase-locking value and a lagged correlation measure (Fig.~\ref{PLV_comparison}, Fig.~\ref{XCorrLag_comparison}). We also tested variants of the MLM in which we partialed out the effect of Euclidean distance from the search information matrix and repeated the single-predictor MLM analysis using the residuals (Fig.~\ref{regressDist}); and in which we substituted the current gene expression correlation matrix with one constructed from genes shown to be predictive of BOLD FC in a previous study (Fig.~\ref{richiardiGenes}; Table.~\ref{tables1}); and in which we substituted the current search information matrix with one estimated from a second independent structual connectivity dataset (Fig.~\ref{differentSearchInformationMatrices}; Table.~\ref{tables3}). Finally, we fit models using a restricted subset of observations, namely the connections that were observed in all seven bands (Table.~\ref{tables4}). Details concerning these additional analyses are included in the \textbf{Supplementary Materials}.

\section*{Discussion}

In this report we propose a technique for estimating whole-brain functional connectivity from ECoG recordings aggregated across multiple subjects. This approach facilitated the construction of (near) whole-brain, band-limited ECoG networks that parsimoniously represented the functional interactions between cortical areas as measured by covariation in regional estimates of sensor signals. Visually, these networks displayed similar topological properties to that observed in BOLD fMRI resting state networks, an observation that we confirmed statistically to be particularly salient in the slowest frequency bands, suggesting an electrophysiological mechanism underpinning inter-regional covariation in BOLD time series. This analysis was complemented by additional multi-modal, multi-linear modeling in which we predicted the magnitude of interregional ECoG FC based on the brain's structural connectivity, its embedding in three-dimensional space, and correlations among brain regions' gene expression profiles. We found that the optimal models included multiple predictors and were able to explain nearly half of the total variance in ECoG FC weights. Moreover, the models displayed utility in predicting single-subject FC patterns but, nonetheless, exhibited subject-specific variation, indicating that they were highly generalizable but also bore the ``fingerprint'' of an individual.

\subsection*{ECoG network architecture and its drivers.}

Our study builds on recent work applying network analysis to study inter-electrode ECoG FC patterns \cite{ortega2008complex, kramer2010coalescence, chu2012emergence, khambhati2016recurring, wilke2011graph, burns2014network, khambhati2016virtual, proix2017individual, chapeton2017stable, solomonwidespread}. Whereas these past studies focused on networks where nodes represented electrodes, which are not consistent across subjects nor do they cover the whole brain, we studied interregional ECoG networks. Our effort is similar in this capacity to another recent paper \cite{solomonwidespread}. Unlike that paper, which aimed in part to relate interregional ECoG FC to cognitive measures, our focus was on characterizing the basic topological principles of ECoG FC organization and predicting connectivity patterns using simple models. Our approach is in line with other predictive models of FC \cite{goni2014resting}, though it has the distinct advantage of predicting FC derived from ECoG, which has clearer neural provenance \cite{dringenberg1998involvement} and is less influenced by motion and physiological artifacts than the BOLD signal \cite{power2012spurious}. The fact that our results show strong similarity between BOLD and ECoG FC, and demonstrate the critical roles of structure, geometry, and genetics in shaping ECoG FC, should dispel some of the concerns surrounding the artifactual nature of BOLD FC.

\subsection*{Cross-modal topological signatures of brain function.}

One of the hallmarks of brain networks is their structural, functional, and cross-modal modular organization \cite{sporns2016modular}. Modules are thought to be critical for both development and evolution by compartmentalizing brain areas that perform similar functions \cite{clune2013evolutionary}. Much emphasis, of late, has been placed on modules in BOLD FC networks, whose boundaries overlap with known cognitive systems, suggesting a possible network-level correlate of psychological and cognitive processes \cite{power2011functional}. Here, we demonstrated that inter-regional ECoG FC networks also exhibit modular architecture, though the overlap with cognitive systems is inexact, a finding that is in line with past studies based on scalp EEG and MEG \cite{mantini2007electrophysiological, marzetti2013frequency}. An important question, then, is why the modules appear different. One possibility is that the ECoG signal carries unique information about patterns of coupling among neuronal populations. Compared to the BOLD signal, ECoG represents a more direct measure of neural activity and with increased temporal resolution it can resolve in greater detail the boundaries of putative modules. More broadly, this mismatch re-emphasizes the brain's multiplex organization, in which brain areas are linked to one another via different connection modalities (e.g., structure, correlated activity, or gene expression) \cite{battiston2017multilayer}.

\subsection*{Basic and clinical utility of prediction.} 

In addition to identifying factors underpinning ECoG FC, the predictive modeling framework has other advantages. Specifically, it makes predictions about the magnitude of ECoG FC between brain regions for which we have no data, complementing previous efforts developing methods to predict missing data in structural connectomes \cite{hinne2017missing} and biomarker data in clinical populations \cite{lo2012predicting}. This is a particularly useful feature for a neuroimaging technique whose coverage is inversely related to the patient's safety: greater coverage is associated with greater risk for inflammation and infection \cite{henle2011first}. Moreover, the prediction goes beyond abstract topological predictors of missing data in complex networks \cite{lu2015toward,pan2016predicting} by incorporating actual physiological constraints in gene and geometry. While an important methodological contribution, these predictions also have potential clinical utility in predicting neurosurgical outcomes. For example, one can simulate the effect of cortical resection by selectively ``lesioning'' structural connections \cite{khambhati2016virtual}, introducing changes in the search information matrix and resulting in an updated ECoG FC prediction. The new and original predictions can be compared to identify connections whose ECoG FC magnitude is expected to increase or decrease as a consequence of the lesion. This prediction, in turn, can be used as a biomarker to guide surgeries, offering an additional quantitative statistic that can be linked to surgical outcomes.

\subsection*{The functional organization of cerebral cortex.}

Understanding the principles that guide the functional organization of neural systems remains a major neuroscientific goal. Towards this end, we identified a set of structural, geometric, and genetic factors that, collectively, predicted the correlation magnitude of electrical activity recorded from distant brain areas. Our findings suggest that the brain's spatial layout and large-scale structural connectivity have especially strong predictive capacity (and presumably) play important roles in determining whether the activity of two brain regions is likely to become coupled. This is in agreement with studies reporting distance-dependent variation of functional connections \cite{margulies2016situating} and close (but not exact) correspondence of interregional correlation to the topology of the underlying structural network \cite{honey2007network, adachi2011functional, hermundstad2013structural, betzel2013multi}.

Interestingly, we found that gene expression correlations had the least predictive capacity of all three factors. That interregional correlations are related, in any way, to the expression levels of specific genes and transcripts is a relatively recent finding \cite{richiardi2015correlated}, and the mechanisms by which these genetic factors can enhance or supress the synchrony of neural activity is not well understood. One possibility is that, like gene-gene coexpression networks in which genes are connected to one another if their expression levels are correlated across samples, interregional correlations of gene expression profiles are driven by sets of functionally-related genes \cite{richiardi2015correlated, rubinov2015wiring}. Allowing for speculation, these groups of genes might perform similar functions, such as ion channel regulation, thereby shaping electrophysiological activity at a low level \cite{gaiteri2014beyond}. Indeed, studies of gene polymorphisms and variants and their role in disease have reported differences in seed-based functional connectivity (estimated from fMRI BOLD) between groups \cite{pezawas20055,meyer2006neural}.

Another possibility is that cytoarchitectural and morphological patterning, both of which influence large-scale structural \cite{goulas2016cytoarchitectonic} and (BOLD) functional connectivity \cite{glasser2016multi}, are genetically regulated \cite{baare2001quantitative}, and thereby have the capacity to influence correlated interregional electrical activity. Genetic regulation of structural covariance matrices has been reported over the course of development \cite{whitaker2016adolescence} and differential gene expression across the adult human cerebral cortex reflects the spatial distribution of cell types \cite{hawrylycz2012anatomically}. In the present study, we did not include an estimate of structural covariance in our predictive model and, to our knowledge, a quantitative large-scale map of cortical cytoarchitecture is unavailable. Future studies should work towards addressing these shortcomings.

\subsection*{Methodological considerations.}

Despite its utility, the predictive framework we develop is correlative in nature \cite{hagmann2008mapping, goni2014resting, richiardi2015correlated}. On the other hand, the spontaneous activity of neural elements (and by extension, FC) arises from their interactions with one another, which serve to constrain some of the observed neurophysiological dynamics \cite{deco2011emerging}. A truly mechanistic model, then, is one that incorporates structure and dynamics to generate synthetic neural activity, which can then be compared to observed activity and its FC \cite{honey2009predicting, adachi2011functional}. Future work could be directed into incorporating both distance-dependence and gene expression levels into mechanistic models. 

The data we analyzed (ECoG recordings and each of the predictors) also represented potential limitations. Despite aggregating recordings from many subjects, there were nonetheless pairs of brain regions for which we had no estimate of ECoG FC. This shortcoming could be addressed in the short term, for example by defining larger brain regions, and in the long term with increased cohort size. In addition, the correlation matrices of brain regions' gene expression profiles are limited in that they were estimated from two subjects worth of observations \cite{jones2009allen,sunkin2013allen}. It is therefore unclear to what extent such matrices are, in fact, representative of the average individual. There are also limitations associated with the calculation of search information, which is based on a structural network of inter-regional, white-matter fiber bundles reconstructed from diffusion-weighted images. The reconstruction procedure is, however, susceptible to false positives and negatives \cite{thomas2014anatomical, reveley2015superficial}. While our use of a consistency-based, group-representative set of tracts reduces this uncertainty, advances in imaging and reconstruction techniques are necessary to mitigate its effect.

\subsection*{Conclusion}

In summary, we present a novel methodological framework for aggregating single-subject ECoG FC into a cohesive, whole-brain network. Our work opens the door for future studies to move beyond inter-electrode networks and investigate properties of inter-regional functional connectivity in ECoG, ultimately documenting how it is modulated with cognitive state and altered in disease. We further show that ECoG FC may be underpinned by a combination of structural, geometric, and genetic factors, and that the contributions made by these factors are relatively consistent across individuals, suggesting a common set of organizational principles.

\section*{Materials and methods}

\subsection*{Functional network reconstruction}

\subsubsection*{Subject-specific, inter-electrode ECoG FC}

We analyzed ECoG recorded from 86 subjects (83 of which had usable data) performing multiple trials of a ``free recall'' experiment (mean$\pm$standard deviation number of trials $=41.9 \pm 25.6$). In this experiment, subjects were presented with a list of words and were later asked to recall as many as possible from the original list. We sought to emulate the ``resting-state'' paradigm common in fMRI BOLD experiments, in which spontaneous neural activity is recorded in the absence of any explicit task commands. The resulting recordings are thought to reflect the brain's intrinsic functional network organization rather than task-related modulations \cite{park2013structural}. To find periods of task-free recordings, we focused on the periods between trials when subjects were awaiting the start of a new trial but otherwise not being asked to perform a task. We extracted 10-second ECoG recordings (epochs) prior to the beginning of each trial. All ECoG data were resampled to 512 Hz. Artifactual channels were discarded, and the remaining channels were average-referenced, stop-filtered to remove line noise and its harmonics, pre-whitened, and bandpass filtered into canonical frequency bands: 1-4 Hz; 4-8 Hz; 8-13 Hz; 13-25 Hz; 25-45 Hz; 85-115 Hz; 140-165 Hz. To reduce boundary effects, we discarded 2.5 and 5.0 seconds of data from the beginning and end of each window and analyzed the approximately stationary middle 2.5 seconds. For each subject and for each trial, we computed inter-electrode FC as a zero-lag Pearson correlation \cite{he2008electrophysiological, kramer2009network, owen2017towards}. Note that we explore other FC measures in \textbf{Supplementary Material: Alternative measures of functional connectivity}. Pairs of electrodes whose correlation magnitude was inconsistent across trials (interquartile range included a value of zero) we excluded from subsequent analyses, focusing instead on correlations that maintained consistent sign and therefore were more likely to be representative of the brain's intrinsic functional architecture rather than task-induced fluctuations. All data are available upon request (\url{http://memory.psych.upenn.edu/RAM_Public_Data}).

\subsubsection*{Mapping electrodes to cortical surface}

Electrode locations were manually digitized using OsiriX software \cite{rosset2004osirix} and stored as voxels in each subject's native coordinate space. These locations were subsequently mapped to the MNI standard coordinate system using the FSL function \verb|img2stdcoord|. We compared each electrode's location in MNI space to vertices on the \emph{fsaverage} pial surface, and assigned each vertex to an electrode if the Euclidean distance between the two was $\le d$ mm. In the main text, we focus on the case where $d = 3$ mm, but we explore $d = 1, 2, 4, 5$ in the \textbf{Supplementary Material}. Each surface vertex was also assigned to one of $N = 114$ brain regions according to a common parcellation-based atlas \cite{cammoun2012mapping}, thereby making it possible to map electrodes to brain regions.

\subsubsection*{Group-representative, inter-regional ECoG FC}

For every pair of brain regions, $i$ and $j$, and for each subject independently, we identified all electrode pairs, $u$ and $v$, where electrode $u$ was assigned to region $i$ and electrode $v$ was assigned to region $j$, and we estimated their average connection weights, generating a subject-specific inter-regional ECoG FC matrix. We estimated the connection weight, $A_{ij}^{\text{ECoG}}$, in the group-representative matrix as the average connection weight over all subjects. We repeated this procedure for each of the seven frequency bands, resulting in band-limited, whole-brain, inter-regional ECoG FC matrices.

\subsection*{Group-representative, inter-regional correlation matrix of gene expression profiles}

The correlation matrix of brain regions' gene expression profiles was reconstructed using a similar approach. We downloaded normalized microarray data from the Allen Brain Institute (\url{http://human.brain-map.org/static/download}) \cite{jones2009allen, hawrylycz2012anatomically, sunkin2013allen}. The full dataset includes six donor brains (aged 18 to 68 years) for which spatially-mapped microarray data were obtained ($\approx$ 60,000 RNA probes). We focused on donors \verb|10021| and \verb|9861| which included samples (893 and 946 sites, respectively) from both the left and right hemispheres. Subsequently, we retained only those samples in the cerebral cortex. Next, we extracted expression profiles for each sample, averaged over duplicate genes, and standardized expression levels across samples as $z$-scores. The standardized measure of any sample, then, measured to what extent a particular gene was differentially expressed at that cortical location relative to the other cortical locations in both hemispheres. 

In addition to microarray data, the Allen Brain Institute also provided coordinates representing the location in MNI coordinates where each sample was collected. This facilitated the mapping of sample sites to brain regions in a procedure exactly analogous to our approach for mapping ECoG electrodes. As a result, we obtained representative expression profiles for each brain region (provided there were nearby samples). For each of the two donor brains, we calculated the region-by-region correlation matrix of standarized expression profiles. Due to the overall density of the whole-brain sampling, we were able to generate an estimate of gene expression correlation (a measure of similarity) for 6286 of 6441 possible region pairs ($\approx$97.6\%).

Note that in the absence of a specific hypothesis about which genes were of particular relevance, we included all genes in our construction of the initial correlation matrices. In the supplement, we follow \cite{richiardi2015correlated} and construct correlation matrices using the same procedures as those described above, but focusing on subsets of genes identified in that paper. For our procedures related to identifying the set of genes that optimized the prediction of ECoG FC, see the later section of this Methods section.

\subsection*{DSI connectome data}

We analyzed a group-representative, whole-brain structural connectivity network -- i.e. a connectome -- generated by combining single-subject data from a cohort of 30 healthy adult participants. Each participant's network was reconstructed from diffusion spectrum images (DSI) in conjunction with state-of-the-art tractography algorithms to estimate the location and strength of large-scale interregional white-matter pathways. Study procedures were approved by the Institutional Review Board of the University of Pennsylvania, and all participants provided informed consent in writing. Details of the acquisition and reconstruction have been described elsewhere \cite{betzel2016optimally, betzel2017modular, betzel2017diversity}. We studied a division of the brain into $N=114$ cortical regions \cite{cammoun2012mapping}. Based on this division, we constructed for each individual an undirected and weighted connectivity matrix, $A \in \mathbb{R}^{N \times N}$, whose edge weights were equal to the number of streamlines detected between regions $i$ and $j$ normalized by the geometric mean of their volumes: $A_{ij} = \frac{S_{ij}}{\sqrt{(V_i V_j)}}$.

\noindent The resulting network was undirected (i.e. $A_{ij} = A_{ji}$). These individual-level networks were then aggregated to form a group-representative network. This procedure can be viewed as a distance-dependent consistency thresholding of connectome data and the details have been described elsewhere \cite{mivsic2015cooperative, betzel2017modular}. The resulting group-representative network has the same number of binary connections as the average individual and the same edge length distribution. This type of non-uniform consistency thresholding has been shown to be superior to other, more commonly used forms \cite{roberts2017consistency}.

\subsection*{fMRI BOLD data}
fMRI BOLD images were acquired during the same scanning session as the DSI data on a 3.0T Siemens Tim Trio whole-body scanner with a whole-head elliptical coil by means of a single-shot gradient-echo T2* ($TR = 1500$ ms; $TE = 30$ ms; $\text{flip angle} = 60^\circ$; $FOV = 19.2$ cm, resolution 3mm x 3mm x 3mm). Preprocessing was performed using FEAT v. 6.0 (fMRI Expert Analysis Tool) \cite{jenkinson2012fsl}. Images underwent the following pre-processing steps: skull-stripping with BET, motion correction with MCFLIRT (FMRIB’s Linear Image Registration Tool; \cite{jenkinson2012fsl}), slice timing correction (interleaved), spatial smoothing with a 6-mm 3D Gaussian kernel, and high pass temporal filtering to reduce low frequency artifacts. We also performed EPI unwarping with fieldmaps to improve subject registration to standard space. Images were transformed to a standard template using FSL's affine registration tool, FLIRT \cite{jenkinson2012fsl}. Subject-specific images were co-registered to their corresponding anatomical images with Boundary Based Registration (BBR \cite{greve2009accurate}) and subsequently registered to the standard MNI-152 structural template via a 12-parameter linear transformation. Lastly, participants' individual anatomical images were segmented into grey matter, white matter, and CSF using the binary segmentation function of FAST v. 4.0 (FMRIB’s Automated Segmentation Tool \cite{zhang2001segmentation}). White matter and CSF masks for each participant were then transformed to native functional space and average timeseries extracted. Images were spatially smoothed using a kernel with a full-width at half-maximum of 6 mm. These values were used as confound regressors on our time series along with 18 translation and rotation parameters as estimated by MCFLIRT \cite{jenkinson2002improved}.

The average time course for each of the 114 cortical regions was extracted and whole-brain inter-regional BOLD FC was computed as the Pearson correlation among all region pairs. The full matrix was subsequently averaged across all subjects to obtain a group-representative estimate (though this averaging procedure can sometimes introduce unwanted biases at the group level \cite{simpson2012exponential}). We denote this BOLD FC matrix as $A^{\text{BOLD}}$.

\subsection*{Network statistics}

\subsubsection*{Modularity maximization}

Real-world networks can be partitioned into node-level clusters called communities by selecting the cluster assignments that optimize a particular objective function. The most popular class are \emph{modularity functions}, which measure the total within-community weight of connections minus what would be expected by chance \cite{newman2004finding}. Maximizing modularity, which results in an estimate of network communities, begins by first defining a modularity matrix, $B$, whose elements are given by $B_{ij} = A_{ij} - P_{ij}$, where $A_{ij}$ and $P_{ij}$ are -- respectively -- the observed and expected weights between nodes $i$ and $j$. Given $B$ and a classification of each node into one of $K$ communities, $\sigma_i \in \{1 , \ldots, K\}$, we can define modularity to be:

\begin{equation}
Q = \sum_{ij} B_{ij} \delta (\sigma_i \sigma_j).
\end{equation}

\noindent Maximizing modularity is accomplished by assigning nodes to communities so that as many positive elements of $B$ fall within communities as possible.

Here, we follow recent work \cite{bazzi2016community} and define our null model to be a constrant free parameter, $\gamma$, so that $B_{ij} = A_{ij}^{\text{ECoG}} - \gamma$. We also set all elements of $B$ for which we have no connectivity data to be zero. This means that unknown or inconsistent connection weights cannot increase or decrease the objective function, $Q$, and therefore have minimal influence on detected communities. We sampled $\gamma$ at 51 linearly-spaced points over the range $[0,0.5]$ and optimize $Q$ using a Louvain-like locally greedy algorithm (100 restarts) \cite{blondel2008fast, jutla2011generalized}. At each value of $\gamma$, we use a consensus clustering algorithm to generate a representative partition from the possibly dissimilar outputs of the maximization algorithm \cite{lancichinetti2012consensus}.

\subsubsection*{Partition similarity}

We assessed the similarity of the partitions uncovered using modularity maximization with the pre-defined system labels \cite{yeo2011organization} using the $z$-score of the Rand index \cite{traud2011comparing}. This measures captures the extent to which two partitions are similar to one another given the number and size of their clusters. For two partitions, $X$ and $Y$, their similarity is calculated as:

\begin{equation}
Z_{XY} = \frac{1}{\sigma_{w_{XY}}}w_{XY} - \frac{W_X W_Y}{W}.
\end{equation}

\noindent Here, $W$ is the total number of node pairs (brain regions) in the network, $W_X$ and $W_Y$ are the number of pairs in the same modules in partitions $X$ and $Y$, respectively, $w_{XY}$ is the number of pairs assigned to the same module in \emph{both} $X$ and $Y$, and $\sigma_{w_{XY}}$ is the standard deviation of $w_{XY}$.

\subsubsection*{Search information}

Anatomical connectivity matrices obtained from diffusion imaging data and reconstructed using deterministic tractography are usually sparse, meaning that only a fraction of all possible connections exist \cite{iturria2008studying, hagmann2008mapping}. Rather than use the sparse connectivity matrix to predict ECoG FC, we generated a full matrix, $S$, whose element $S_{ij}$ indicates the information (in bits) required to follow the shortest path from node $i$ to node $j$ \cite{rosvall2005searchability}. Let $\pi_{s \rightarrow t} = \{ A_{si}, A_{ij}, \ldots , A_{kt} \}$ be the series of structural edges that are traversed along the shortest path from a source node, $s$, to a different target node, $t$, and $\Omega_{s \rightarrow t} = \{s,i,j,\ldots,k,t\}$ be the sequence of nodes along the same path. The probability of following this path under random walk dynamics is given by $P(\pi_{s \rightarrow t}) = \prod_{i \in \Omega^*_{s \rightarrow t}} \frac{\pi_{i \rightarrow t}^{(1)}}{s_i}$, where $s_i = \sum_j A_{ij}$ is the weighted degree of node $i$, $\pi_{i \rightarrow t}^{(1)}$ is the first edge on the shortest path from $i$ to $t$ and $\Omega^*_{s \rightarrow t} = \{s,i,j,\ldots,k\}$ is the shortest path node sequence excluding the target node. The amount of information (in bits) required to access this shortest path, then, is given by $S(\pi_{s \rightarrow t}) = \log_2(P(\pi_{s \rightarrow t}))$.

We can treat every pair of nodes $\{i,j\}$ as the source and target, respectively, and (provided that there exists a unique shortest-path from $i$ to $j$) we can compute $S(\pi_{i \rightarrow j})$ for all such pairs. The resulting matrix, $S$, termed ``search information'', has been shown to be a good predictor of BOLD FC \cite{goni2014resting} and may be modulated in certain neurological disorders \cite{wirsich2016whole}.

\subsubsection*{Network null model}

We counted the number of jointly strong and long connections for ECoG FC networks that represented different frequency bands. In Figs.~\ref{fcProperties}g,h we compared those counts across frequency bands. To demonstrate the statistical significance of these findings, we also compared counts for random networks generated under a particular null model. This null model preserved the binary topology and spatial embedding of each frequency-specific network, but otherwise scrambled edge weights across frequencies. Given a pair of nodes $i$ and $j$ whose connection weights across frequency bands are specified by $A_{ij}^f$, where $f = \{1 , \ldots , 7\}$, we generated random networks by randomly permuting the order of those weights across frequencies and repeating this process for all pairs of nodes. It was sometimes the case that for certain pairs of nodes a connection was only observed in a subset of frequencies. In this event, the permutation was only carried out over those frequency bands in which the connection was observed.

\subsection*{Materials and Methods: Gene-ECoG Optimization}

In the main text we briefly describe a procedure for identifying genes that are related to ECoG FC. In general, we sought the list of $K$ genes, $\Gamma^{K} = \{g_1,\ldots,g_K \}$ whose brain-wide co-expression matrix was maximally correlated with ECoG FC. While the exact solution of this optimization problem is computationally intractable (the full list included 29130 genes), we could define an objective function and use numerical methods to obtain an approximate solution.

The objective function we sought to minimize was defined as follows. Let $G_1(\Gamma)$ and $G_2(\Gamma)$ be the gene co-expression matrices for each of the two donor brains calculated using the gene list, $\Gamma$. We can then vectorize each matrix by extracting its upper triangle of non-zero elements and, after doing the same for the ECoG FC matrix, $\mathbf{A}^{\text{ECoG}}$, calculate the correlation of gene expression with ECoG FC, resulting in two correlation coefficients $\rho_1$ and $\rho_2$. In general, we want the magnitudes of $\rho_1$ and $\rho_2$ to be as large as possible. Accordingly, we defined our objective function to be $F(\rho_1, \rho_2) = \min(\rho_1,\rho_2)$, so that correspondence of any gene list, $\Gamma$, with ECoG FC is only as good as the worse of the two donor brains correlations.

As noted earlier, optimizing this function is computationaly intractable, so we used a simulated annealing algorithm to generate estimates of the solution. In general, simulated annealing works by proposing initial estimates of the solution (that are usually poor), making small changes to these estimates and evaluating whether or not these changes improve the estimate. The algorithm begins in a ``high temperature'' phase, during which even changes that result in inferior estimates can be accepted, making it possible to explore the landscape of possible solutions. Gradually, a temperature parameter is reduced so that in later phases only solutions that result in improvements are accepted.

In our case, the algorithm was initialized with a temperature of $t_0 = 2.5$ and a randomly-generated list of $K$ genes, $\Gamma$, which represented our initial estimate of the solution. From this list we constructed matrices $G_1(\Gamma)$ and $G_2(\Gamma)$, calculated $\rho_1$ and $\rho_2$, and then evaluated the objective function, $F(\rho_1,\rho_2)$. With each iteration, the temperature was reduced slightly, ($t_i = t_{i - 1} \times 0.99975$) and one gene randomly selected from $\Gamma$ was replaced with a novel gene. We then used this new list, $\Gamma^{\prime}$, to construct $G_1(\Gamma)^{\prime}$ and $G_2(\Gamma)^{\prime}$, from which we eventually obtained a new value of the objective function, $F(\rho_1^{\prime},\rho_2^{\prime})$. If $F(\rho_1^{\prime},\rho_2^{\prime}) > F(\rho_1,\rho_2)$, then we replaced $\Gamma$ with $\Gamma^{\prime}$ and the algorithm proceeded to the next iteration. Otherwise, we accepted the $\Gamma^{\prime}$ with probability $\exp(-\frac{[F(\rho_1,\rho_2)-F(\rho_1^{\prime},\rho_2)^{\prime}]}{t_i})$, where $t_i$ is the temperature at the current iteration. The algorithm continued for either 200000 total iterations or 10000 consecutive iterations with no change in $\Gamma$.

The result of simulated annealing will usually vary somewhat from run to run. Accordingly, we repeated the algorithm 50 times. We also varied the number of genes, $K$, from 10 to 360 in increments of 10. We chose the optimal $K$ to be the value at which the objective function was on average greatest over the 50 repetitions. Rather than treat any of the 50 estimated solutions as representative, we calculated how frequently each gene appeared across the ensemble of all 50, and we compared this frequency to what we would expect in 50 samples of $K$ genes. We retained only those genes that appeared more frequently than expected (false discovery rate controlled at $q = 0.05$). These genes represented the ``optimized list'' and were submit to the ontology analysis.

\section*{Acknowledgements}
The resting state fMRI and diffusion imaging data collection was funded by an Army Research Office through contract number W911NF-14-1-0679 awarded to DSB. RFB, JS, and DSB acknowledge support from the John D. and Catherine T. MacArthur Foundation, the Alfred P. Sloan Foundation, the National Institute of Health (2-R01-DC-009209-11, 1R01HD086888-01, R01-MH107235, R01-MH107703, R01MH109520, 1R01NS099348 and R21-M MH-106799), the Office of Naval Research, and the National Science Foundation (BCS-1441502, CAREER PHY-1554488, BCS-1631550, and CNS-1626008). AEK and DSB acknowledge support from the Army Research Laboratory through contract number W911NF-10-2-0022. JDM acknowledges support from the National Institute of Health, award 1-DP5-OD021352. The content is solely the responsibility of the authors and does not necessarily represent the official views of any of the funding agencies. We also thank Professor Michael Kahana and the group associated with the DARPA Restoring Active Memory program, who were responsible for collecting and publicly sharing the ECoG data used in this paper.
%
%
%
%
%
%
%
\clearpage
\beginsupplement

\section*{Supplementary Materials}
In the main text, we describe a procedure for estimating whole-brain inter-regional functional connectivity (FC) from ECoG recordings aggregated across a large cohort of subjects. In this supplement, we document how variation in the parameter governing the spatial resolution with which electrodes are mapped to brain regions influences the topological properties of the reconstructed network. In addition, we explore variations of the MLM procedure that include: (i) an alternative method for constructing the gene expression correlation matrix by excluding certain subsets of genes; (ii) the use of a novel and much higher-resolution diffusion imaging dataset to construct the interregional search information matrix; (iii) enforcing that the model is fit using the same number of observations in each frequency band in order to discount the possibility that variation in model performance across frequencies is driven by disparate numbers of observed connections. Finally, we show that by extending the linear modeling framework to the level of individual brain systems, we gain additional insight into factors that influence system-level network organization of ECoG FC. In general, these supplementary studies support and complement	 the results presented in the main text.

\subsection*{Distance Threshold}

The process of network construction and analysis of the resulting network depends upon two free parameters. The first parameter is the distance threshold used to map ECoG electrodes to vertices on the brain surface mesh. Effectively, this threshold controls the specificity of the electrode-to-region mapping. More specifically, its value determines the radius of a sphere surrounding each electrode; any surface vertex within that sphere is considered to be associated with that electrode.

\begin{figure*}[t]
\begin{center}
	\centerline{\includegraphics[width=0.8\textwidth]{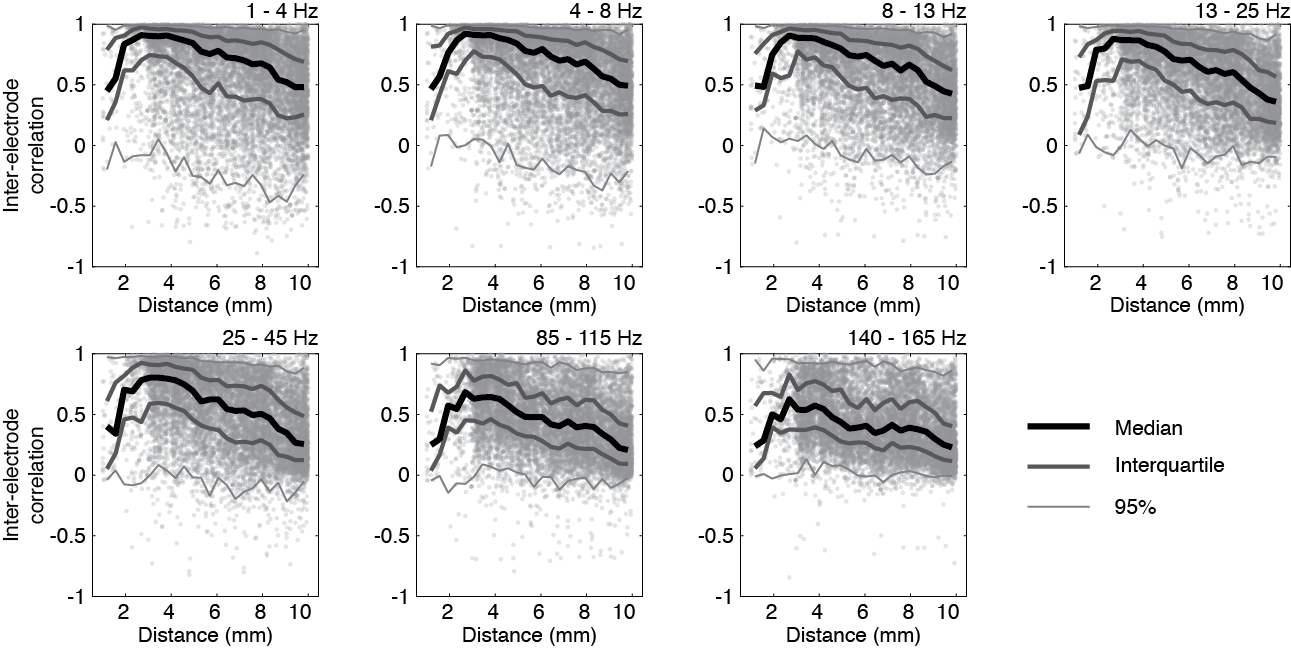}}
	\caption{\textbf{Relationship of inter-electrode distance with inter-electrode FC.} In constructing inter-regional ECoG FC networks, we aggregated inter-electrode correlations based on inter-electrode distances. The mapping of electrodes to brain surface vertices (and subsequently to brain regions) depended upon a distance threshold. While FC magnitude decays as a function of distance, we see that within $\approx$5 mm, the decrease is minimal. This observation motivates selecting distance thresholds of comparable length.} \label{spatialAutocorrelation}
\end{center}
\end{figure*}

This threshold has implications for the organization and interpretation of the networks that we construct. A stringent value (small spheres) results in an electrode-to-region mapping with correspondingly high spatial specificity -- i.e. we can be confident that the ECoG signal recorded from an electrode and the functional connections made by that electrode are localized to one or a small set of brain regions. On the other hand, choosing the parameter that results in the highest spatial specificity may not be ideal. Neurobiologically, we expect nearby points on the brain's surface to display similar activity and be driven by common cortical sources \cite{owen2017towards}. Indeed, the spatial autocorrelation of the processed ECoG time series (at the electrode level) suggests that nearby electrodes display time series that are highly correlated with one another. For example, over a range of 10 mm, the median autocorrelation of 1-4 Hz activity is $r \approx 0.72$ (Fig.~\ref{spatialAutocorrelation}). This value attenuates in higher frequency bands, but even for the fastest band considered here (140-165 Hz) the spatial autocorrelation over the same 10 mm range was $r \approx 0.40$. Accordingly, setting the distance threshold to have high spatial specificity likely does not significantly improve our ability to resolve contributions from unique sources to the ECoG signal. Additionally (and from a more practical perspective), stringent thresholds influence the number of edges (pairs of brain regions) for which we can estimate the magnitude of ECoF FC.  Smaller spheres mean that electrodes are associated with fewer brain regions, which results in correspondingly fewer pairs of brain regions with observed functional connections.

\begin{figure*}[t]
\begin{center}
	\centerline{\includegraphics[width=0.8\textwidth]{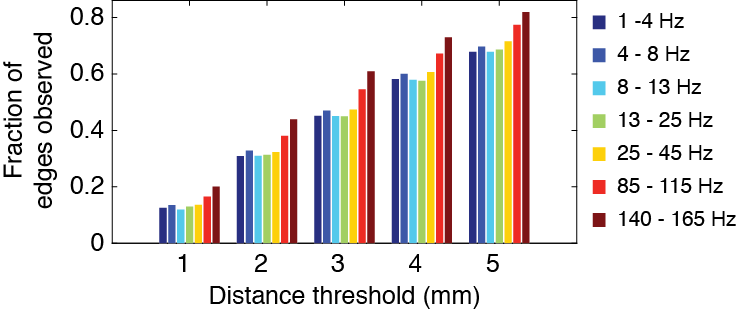}}
	\caption{\textbf{Fraction of connections observed in each frequency band as a function of distance threshold.} Bar colors represent frequency bands. All bands are grouped in septets based on distance threshold.} \label{networkSparsity}
\end{center}
\end{figure*}

\begin{figure*}[t]
\begin{center}
	\centerline{\includegraphics[width=0.8\textwidth]{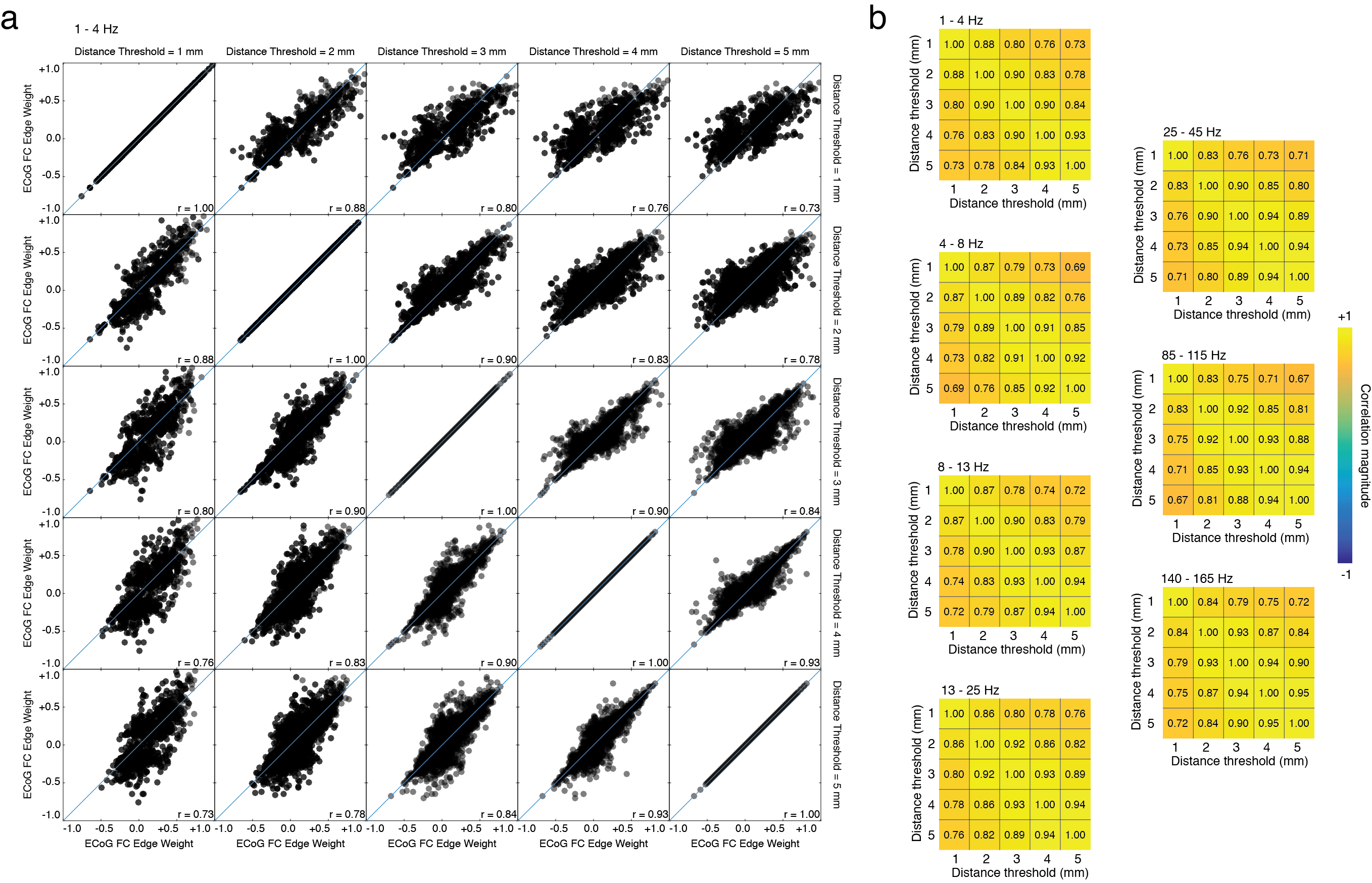}}
	\caption{\textbf{Effect of distance threshold on ECoG FC edge weights.} (\emph{a}) Scatterplots of ECoG FC edge weights for every pair of distance thresholds applied to the 1-4 Hz frequency band. Note that the magnitude correlation is shown in the bottom corner of each subplot. (\emph{b}) We summarize (\emph{a}) for all frequency bands by reporting correlation coefficients alone.} \label{variationOfDistanceThreshold}
\end{center}
\end{figure*}

With some general guidelines but no \emph{a priori} knowledge as to what might be an appropriate distance threshold, we tested thresholds of 1, 2, 3, 4, and 5 mm. The resulting network varied in terms of the number of brain region pairs for which we observed at least one connection across all subjects (Fig.~\ref{networkSparsity}), but overall the structure of the network was highly consistent. We computed the Pearson correlation among the set of mutually observed connections for networks estimated at every pair of thresholds, from 1 - 5 mm (Fig.~\ref{variationOfDistanceThreshold}). In detail, we show the scatterplot for 1-4 Hz (Fig.~\ref{variationOfDistanceThreshold}a). We summarize the results for the other frequency bands by reporting correlation coefficients (Fig.~\ref{variationOfDistanceThreshold}b). The high levels of correlation indicate that the networks we generate with different distance thresholds, while different in terms of the number of observed connections, are surprisingly similar. This finding suggests that the ECoG FC network and its organization are largely robust to the choice of distance threshold. Without a clear motivation to adopt one threshold over another, we focused on a threshold of 3 mm in the main text.

\subsection*{Alternative measures of functional connectivity}

In the main text, we define the functional connection weight between electrodes (and eventually between brain regions) as a zero-lag Pearson correlation of those regions' activity. While this measure of FC has been useful in dealing with the slow fMRI BOLD signal \cite{smith2011network}, other measures have achieved greater popularity when dealing with electrophysiological signals. In this supplementary section, we focus on two other measures: the lagged, normalized cross-correlation \cite{kramer2009network} and phase locking \cite{lachaux1999measuring}. We show that the networks obtained using these measures are, broadly, similar to those we analyzed in the main text.

\subsubsection*{Lagged, normalized cross-correlation}

One criticism of zero-lag correlations is that they do not account for conduction delays and other time-dependent biophysical properties of interregional communication \cite{vicente2008dynamical} (though this is disputed by others, as zero-lag correlations can emerge naturally as a consequence of the underlying structural network organization \cite{gollo2014mechanisms}). One means of circumventing potential issues with zero-lag correlations is to consider peaks in the cross-correlogram at non-zero lags. The lagged and normalized cross-correlation between two time series does precisely this. For two time series, $x_{i}(t)$ and $x_{j}(t)$, it computes a normalized cross-correlation value at different lags, $\tau$, up to a maximum lag of $\tau_{\text{max}}$:

\begin{equation}
\rho_{ij}(\tau) = \frac{1}{\sigma_i \sigma_j (n - 2\tau)}\sum_{t = 1}^{n - \tau} (x_i(t) - \bar{x}_i)(x_j(t + \tau) - \bar{x}_j).
\end{equation}

\noindent In this expression $\sigma_i$ and $\bar{x}_i$ are the sample standard deviation and mean of time series $x_i(t)$, and $n$ is the number of points in the time series. If $x_{i}(t)$ and $x_{j}(t)$ represent electrode recordings, then we can define the FC magnitude between electrodes $i$ and $j$ to be the value $\rho_{ij}(\tau)$ corresponding to the $\tau$ that satisfyies $\max_{\tau} |\rho_{ij}(\tau)|$, where $|\cdot|$ is the absolute value. Rather than considering all possible lag values, we set $\tau_{\text{max}}$ equal to 1 second (512 samples).

\subsubsection*{Phase-locking value}

To this point, the FC measures discussed here and in the main text are based on the presupposition that there exists a linear (temporal) relationship between the activities of two electrodes. Another approach, and one that has a long history in the electrophysiology literature, is based on the phase offsets of electrodes' activity with one another \cite{lachaux1999measuring}. A real-valued time series $x_i(t)$ can be decomposed into a complex, analytic signal \emph{via} the Hilbert transform. This results in two components, one real ($x_i(t)^\text{real}$) and another imaginary ($x_i(t)^\text{imag}$). These components satisfy the relationship $x_i(t) = x_i(t)^\text{real} + j x_i(t)^\text{real}$, where $j = \sqrt{-1}$ is an imaginary number. From these components, we can calculate the instantaneous phase, $\phi_i(t)$, at each time point $t$.

Given two time series of phase angles, $\phi_i(t)$ and $\phi_j(t)$, we compute their relative phase $\theta_{ij}(t) = \phi_i(t) - \phi_j(t)$. After remapping $\theta_{ij}(t)$ to the interval $[-\pi,\pi]$, we compute the time-averaged phase-locking between $i$ and $j$ as:

\begin{equation}
PLV_{ij} = \frac{1}{n}\sum_{t = 1}^n e^{\theta_{ij}}.
\end{equation}

\noindent To demonstrate that both the lagged, normalized cross-correlation and phase-locking return similar estimates of the network as those obtained in the main text, we computed the correlation of their connection weights with the zero-lag connection weights.

\begin{figure*}[t]
\begin{center}
	\centerline{\includegraphics[width=1\textwidth]{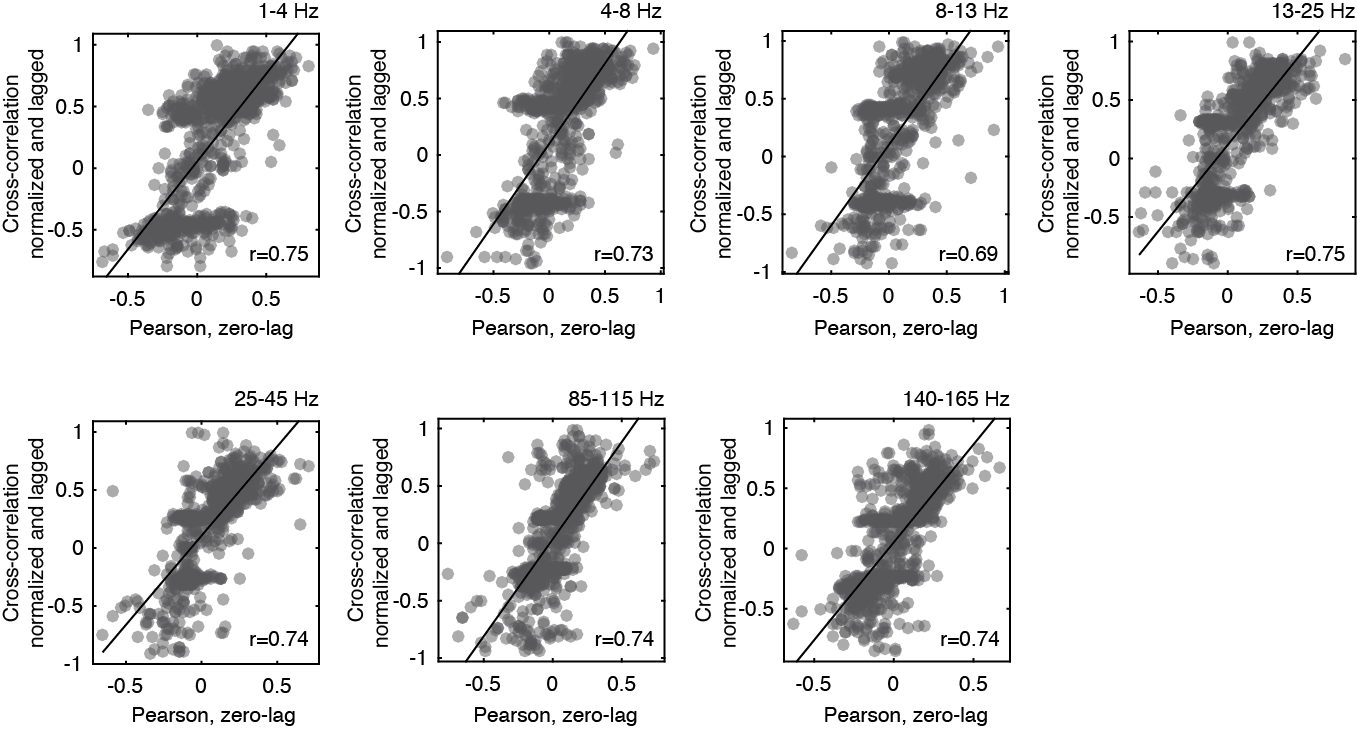}}
	\caption{\textbf{Relationship of lagged and normalized cross-correlation with zero-lag cross correlation.} Each panel shows edge-by-edge scatterplot of edge weights for different frequency bands.} \label{XCorrLag_comparison}
\end{center}
\end{figure*}

\begin{figure*}[t]
\begin{center}
	\centerline{\includegraphics[width=1\textwidth]{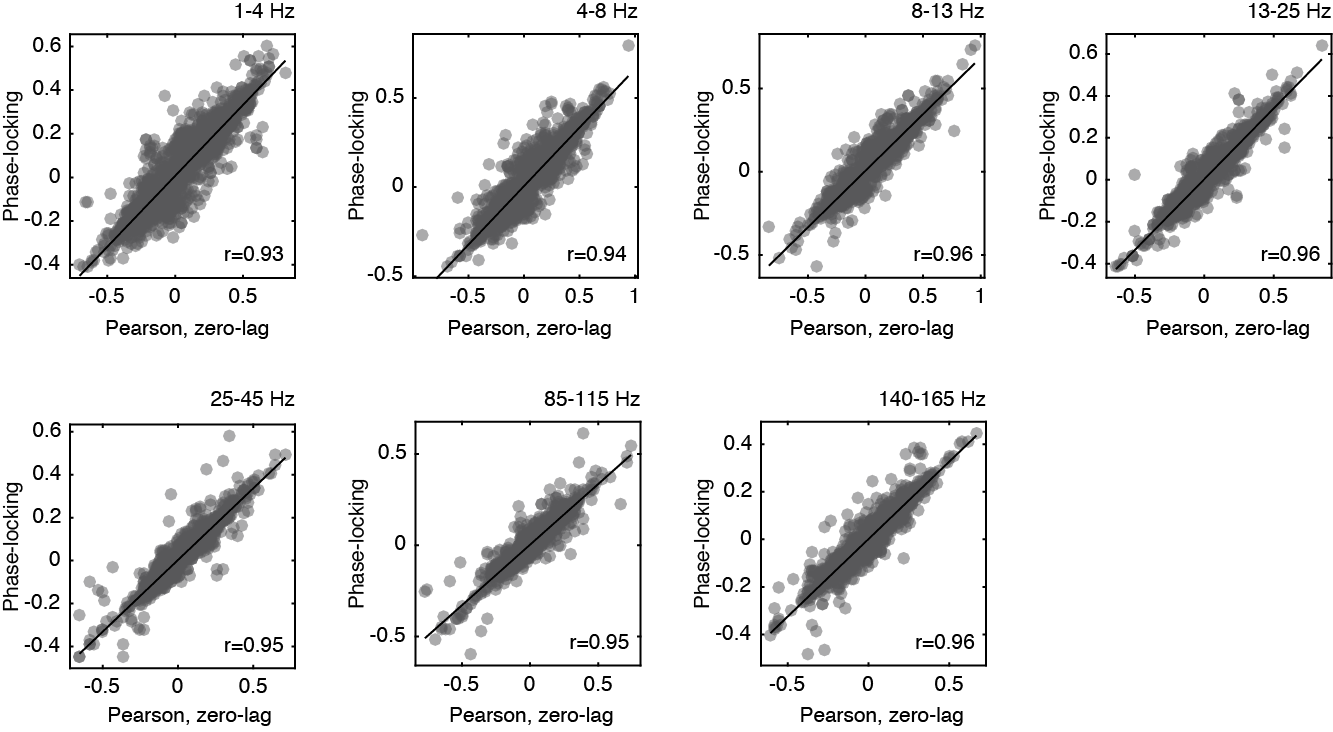}}
	\caption{\textbf{Relationship of phase-locking values with zero-lag cross correlation.} Each panel shows edge-by-edge scatterplot of edge weights for different frequency bands.} \label{PLV_comparison}
\end{center}
\end{figure*}

In general, the ECoG FC matrices for both alternatve FC measures proved highly correlated with the zero-lag correlations analyzed in the main text. In the case of the lagged, normalized cross-correlation, we found that connection weights were consistently correlated across all frequency bands, ranging from $r = 0.69$ to $r = 0.75$ (Fig.~\ref{XCorrLag_comparison}). It is worth noting that connection weights defined using the lagged, normalized cross-correlation measure appeared increasingly bi-modal; for many connections there existed a correlation at a non-zero lag with a magnitude that exceeded the zero-lag correlation (see, for example, the scatter plot for the 1-4 Hz comparison in Fig.~\ref{XCorrLag_comparison}). For the phase-locking measure, the correspondence was even stronger, with correlations ranging from a minimum of $r = 0.93$ to a maximum of $r = 0.96$ (Fig.~\ref{PLV_comparison}). These findings indicate that, despite the simplicity of the connectivity measure used in their construction, the ECoG FC matrices studied in the text are strikingly similar to matrices defined using more complicated measures of connectivity.

It is important to note that there may be systematic subject level differences between the measures (that get ``smoothed out'' when subjects are aggregated together) and it is possible that certain measures are better-suited for particular research questions.
	
\subsection*{Variants of the multi-linear model}

In the main text, we used a multi-linear modeling (MLM) framework to generate predictions of ECoG FC from three predictors: search information (a measure based on structural connectivity), interregional distance, and correlated gene expression profiles. In this section, we show that the results presented in the main text are qualitatively unchanged when we perturb or vary these predictors slightly. Namely, we show that (1) single-predictor models are significantly outperformed by models that include multiple predictors and (2) that our predictions become increasingly accurate in higher frequency bands.

We explore different variants of the MLMs described in the main text. In the first, we test the predictive capacity of search information after partialing out the effect of Euclidean distance. Another of these variants involves redefining the gene expression correlation matrix to take into account specific subsets of genes. Another variant involves taking advantage of recent advances in diffusion imaging sequences and longer scan times to improve our reconstruction of structural connectivity networks using tractography. We use these new structural connectivity matrices to generate an alternative estimate of search information. Finally, the number of ECoG FC connections varied across frequency bands. This variability in the number of observations could introduce potential biases in our ability to accurately predict ECoG FC using the MLM (e.g., if one frequency band includes hard-to-predict connections and another does not). To test and confirm that this was not the case, we repeated the analysis reported in the main text using only connections that were observed in all seven frequency bands.

\subsubsection*{Disentangling the effect of structure from Euclidean distance}

A growing number of studies have demonstrated that Euclidean distance plays a critical role in shaping the organization of brain structural connectivity \cite{chen2006wiring, betzel2016generative, roberts2016contribution}, with some recent studies arguing that connection weights are wholly determined by Euclidean distance or the curvilinear length of fiber tracts \cite{ercsey2013predictive, song2014spatial}.

\begin{figure*}[t]
	\begin{center}
		\centerline{\includegraphics[width=1\textwidth]{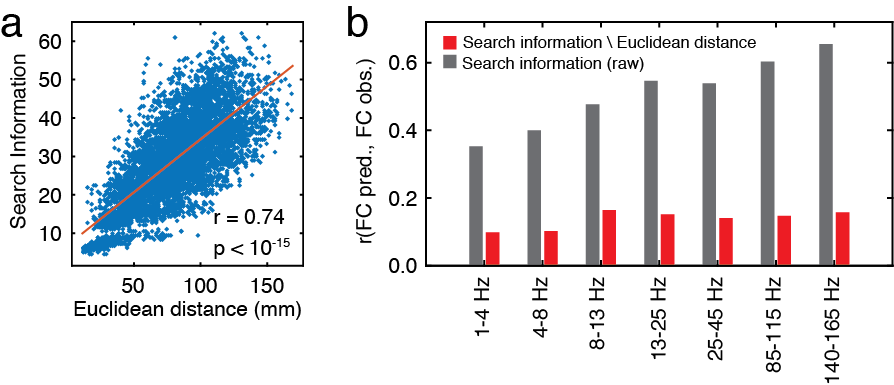}}
		\caption{\textbf{Single-predictor MLM results before and after partialling Euclidean distance from search information.} Each bar represents the correlation of predicted and observed ECoG FC. Gray bars represent correlations obtained using the raw search information matrix and red bars represent correlations obtained using the search information matrix after the effect of Euclidean distance was partialed (regressed) out from its edge weights.} \label{regressDist}
	\end{center}
\end{figure*}

In the main text, we describe a multi-linear model in which we include both Euclidean distance and search information as predictors. Search information, of course, is a measure derived from structural connectivity and, as expected, is co-linear with Euclidean distance (Fig.~\ref{regressDist}a). We demonstrate using single-predictor models that search information has comparable predictive capacity to Euclidean distance. An important test, then, is to determine whether search information can be used to explain variance among ECoG FC beyond that explained by Euclidean distance.

To address this issue, we used regression analysis to partial out the effect of Euclidean distance from the search information matrix. The residuals of this regression analysis are, by definition, uncorrelated with Euclidean distance. We then used the residuals as the predictor in multi-linear models to predict the ECoG FC in all seven frequency bands. If search information explains no additional variance in ECoG FC, we would expect these residuals to be uncorrelated with the ECoG FC matrix.

While the overall correlation magnitude was attenuated (Fig.~\ref{regressDist}b), an effect that has been described in predictive models of fMRI BOLD \cite{goni2014resting}, we nontheless find that the residuals of search information after partialing out Euclidean distance are still correlated with ECoG FC (maximum $p$-value, $p \approx 1.3 \times 10^{-7}$). In summary, this supplemental result demonstrates that, in agreement with past studies \cite{ercsey2013predictive, song2014spatial, betzel2016generative}, Euclidean distance and measures derived from structural connectivity are largely co-linear, but that those measures (in this case search information) are not entirely explained by Euclidean distance and that they are still predictive of ECoG FC.

\subsubsection*{Variants of the gene expression correlation matrix}

\begin{figure*}[t]
\begin{center}
	\centerline{\includegraphics[width=0.75\textwidth]{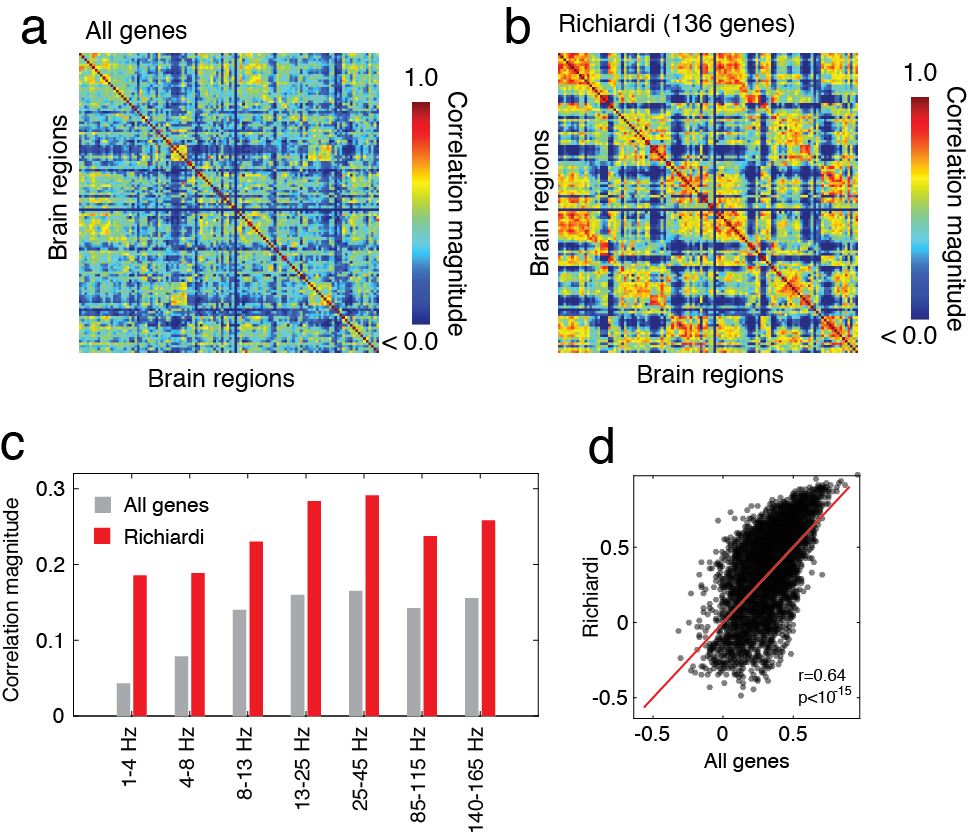}}
	\caption{\textbf{Gene expression profile correlation matrix variants.} (\emph{a}) Co-expression matrix estimated from all genes. (\emph{b}) Co-expression matrix estimated from the 136 genes identified in \cite{richiardi2015correlated}. (\emph{c}) Results of single-factor models using the matrices shown in panels \emph{a} and \emph{b}. (\emph{d}) Scatterplots showing the relationships of elements in each matrix with one another.} \label{richiardiGenes}
\end{center}
\end{figure*}

In the main text, we generated an inter-regional correlation matrix of gene expression profiles from Allen Brain Institute data. Initially we used a set of 29131 genes to construct the matrix. Eventually we performed an optimization procedure to identify much smaller subsets of genes whose matrix of correlated expression profiles was maximally correlated with ECoG FC. Recent work using these same data, however, identified 136 genes associated with fMRI BOLD FC \cite{richiardi2015correlated}. We refer to this gene group as the ``Richiardi'' subset, with corresponding gene co-expression matrix, $\mathbf{G}^R$. In general, $\mathbf{G}^R$ and $\mathbf{G}$ were similar to one another (correlation magnitude of their elements was $r = 0.64$; Fig.~\ref{richiardiGenes}d). To test whether $\mathbf{G}^R$ improved model performance, we compared the single-factor models using $\mathbf{G}$ and $\mathbf{G}^R$ to predict ECoG FC. Overall, $\mathbf{G}^R$ resulted in a substantial improvement (Fig.~\ref{richiardiGenes}c) in all frequency bands, but was still far less of an improvement than what we found using our optimization procedure.

In addition to the single-factor models, we replaced $\mathbf{G}$ with $\mathbf{G}^R$ in all the MLMs, including multi-factor models. We observed consistent improvements in all frequency bands. However, like the single-factor models, these improvements were notably less than those observed when we replaced $\mathbf{G}^R$ with $\mathbf{G}^{opt}$ (Table.~\ref{tables1}).

\subsubsection*{Variant of structural connectivity}

In addition to the diffusion data reported in the main text, which has been analyzed elsewhere \cite{betzel2016optimally, betzel2017diversity, betzel2017modular}, we repeated our analyses using a novel diffusion imaging dataset. This dataset had the advantage of assessing 730 diffusion directions over the course of 53 minutes of multi-band acquisition. Specifically, ten healthy adult human subjects (m) were imaged as part of an ongoing data collection effort at the University of Pennsylvania; the subjects provided informed consent in writing, in accordance with the Institutional Review Board of the University of Pennsylvania. All scans were acquired on a Siemens Magnetom Prisma 3 Tesla scanner with a 64-channel head/neck array at the University of Pennsylvania. Each data acquisition session included both a diffusion spectrum imaging (DSI) scan as well as a high-resolution T1-weighted anatomical scan. The diffusion scan was 730-directional with a maximum $b$-value of 5010s/mm$^{2}$ and TE/TR = 102/4300 ms, which included 21 $b=0$ images. Matrix size was 144$\times$144 with a slice number of 87. Field of view was 260$\times$260mm$^2$ and slice thickness was 1.80mm. Acquisition time per scan was 53:24min, using a multi-band acceleration factor of 3. The anatomical scan was a high-resolution three-dimensional T1-weighted sagittal whole-brain image using a magnetization prepared rapid acquisition gradient-echo (MPRAGE) sequence. It was acquired with TR = 2500 ms; TE=2.18 ms; flip angle = 7 degrees; 208 slices; 0.9mm thickness.

DSI is highly sensitive to subject movement \cite{bilgic2013fast}, which can cause significant distortions in the reconstructed ODFs if not corrected. Motion correction is typically applied by determining an affine or non-linear transform to align each DWI volume to a reference derived from the high-signal $b=0$ images. The high $b$-values used in DSI present a problem for this approach, as the low signal in many of the volumes leads to poor registration. To address this, we interspersed $b=0$ volumes in the scan sequence, one for every 35 volumes. An initial average template was produced by averaging the $b=0$ images together and then improved by registering the $b=0$ images to the initial template and re-averaging. Each $b=0$ was finally re-registered to the improved template, and then each volume in the DSI scan was then motion corrected by applying the transformation calculated for the closest $b=0$ volume. Motion correction also impacts the effective $b$-matrix directions since the rotated images are no longer aligned with the scanner; therefore the transforms applied to motion correct each volume were also used to rotate the corresponding $b$-vectors \cite{leemans2009b}. The processing pipeline was implemented using Nipype \cite{gorgolewski2011nipype} with registration performed using the Advanced Normalization Tools (ANTs)\cite{avants2011reproducible}.

Using DSI-Studio (http://dsi-studio.labsolver.org), orientation density functions (ODFs) within each voxel were reconstructed from the corrected scans using GQI \cite{yeh2010generalized}. We then used the reconstructed ODFs to perform a whole-brain deterministic tractography using the derived QA values in DSI-Studio \cite{yeh2013deterministic}. We generated 1,000,000 streamlines per subject, with a maximum turning angle of 35 degrees \cite{bassett2011conserved} and a maximum length of 500 mm \cite{cieslak2014local}. Networks were then constructed in an identical manner to those described in the Methods section of the main text.

\begin{figure*}[t]
\begin{center}
	\centerline{\includegraphics[width=0.75\textwidth]{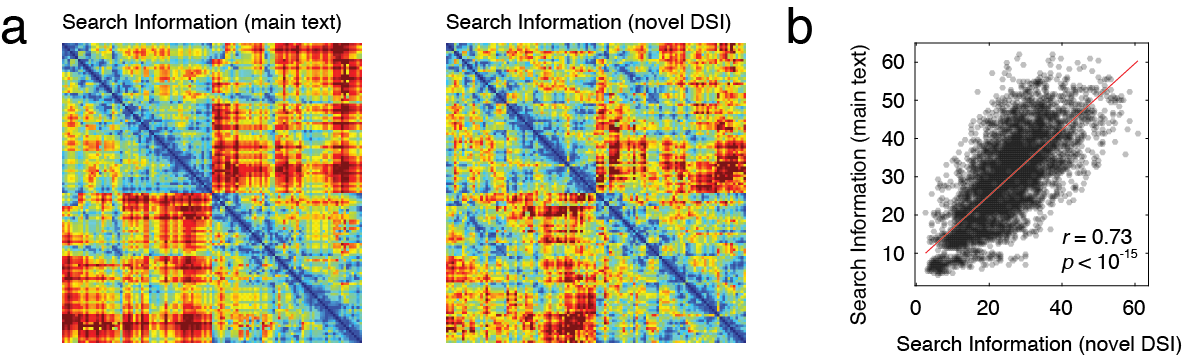}}
	\caption{\textbf{Search information matrix variant.} (\emph{a}) \emph{left}: search information matrix analyzed in the main text. \emph{right}: matrix generated using alternative SC dataset. (\emph{b})  Scatterplots showing the relationships of elements in each matrix with one another.} \label{differentSearchInformationMatrices}
\end{center}
\end{figure*}

From these data, we estimated a group-representative structural connectivity matrix and, subsequently, the search information between every pair of brain regions (Fig.~\ref{differentSearchInformationMatrices}a). This new estimate was highly correlated with the search information matrix presented in the main text ($r = 0.73$, $p < 10^{-15}$; Fig.~\ref{differentSearchInformationMatrices}b). Along with inter-regional distance and gene expression correlation matrices, we included the new search information estimate in place of the one described in the main text and repeated our MLM analysis. The results of this supplementary analysis were similar to those presented in the main text, namely increased model performance as an increasing function of frequency and optimal models for all frequency bands that included multiple predictors (Table.~\ref{tables3}).

\subsubsection*{Same set of edges}

Finally, to address the issue that the variation in model performance was dependent upon the number of connections (observations) in each frequency band, we repeated our MLM analysis on a restricted set of connections. Specifically, we focused on the 1744 connections that were observed across all seven frequency bands. As with the other MLM variants, we found that the optimal models included multiple predictors, with ECoG FC in six of the seven frequency bands best explained by the full model (Table.~\ref{tables4}).

\subsection*{System-level multi-linear models}

In the main text, we aggregated all observed connections and leveraged measures derived from the brain's structural connectivity, spatial embedding, and correlations among brain regions' gene expression profiles to predict the weights of those connections. In this section, we explore the effect of partitioning connections into blocks that fall within and between cognitive systems \cite{yeo2011organization} and fitting models independently to each block. This procedure makes it possible to obtain a more nuanced understanding of the factors that differentially influence connectivity among select systems and also identify systems that are not as well predicted given the factors that we include in our models.

\begin{figure*}[t]
\begin{center}
	\centerline{\includegraphics[width=0.75\textwidth]{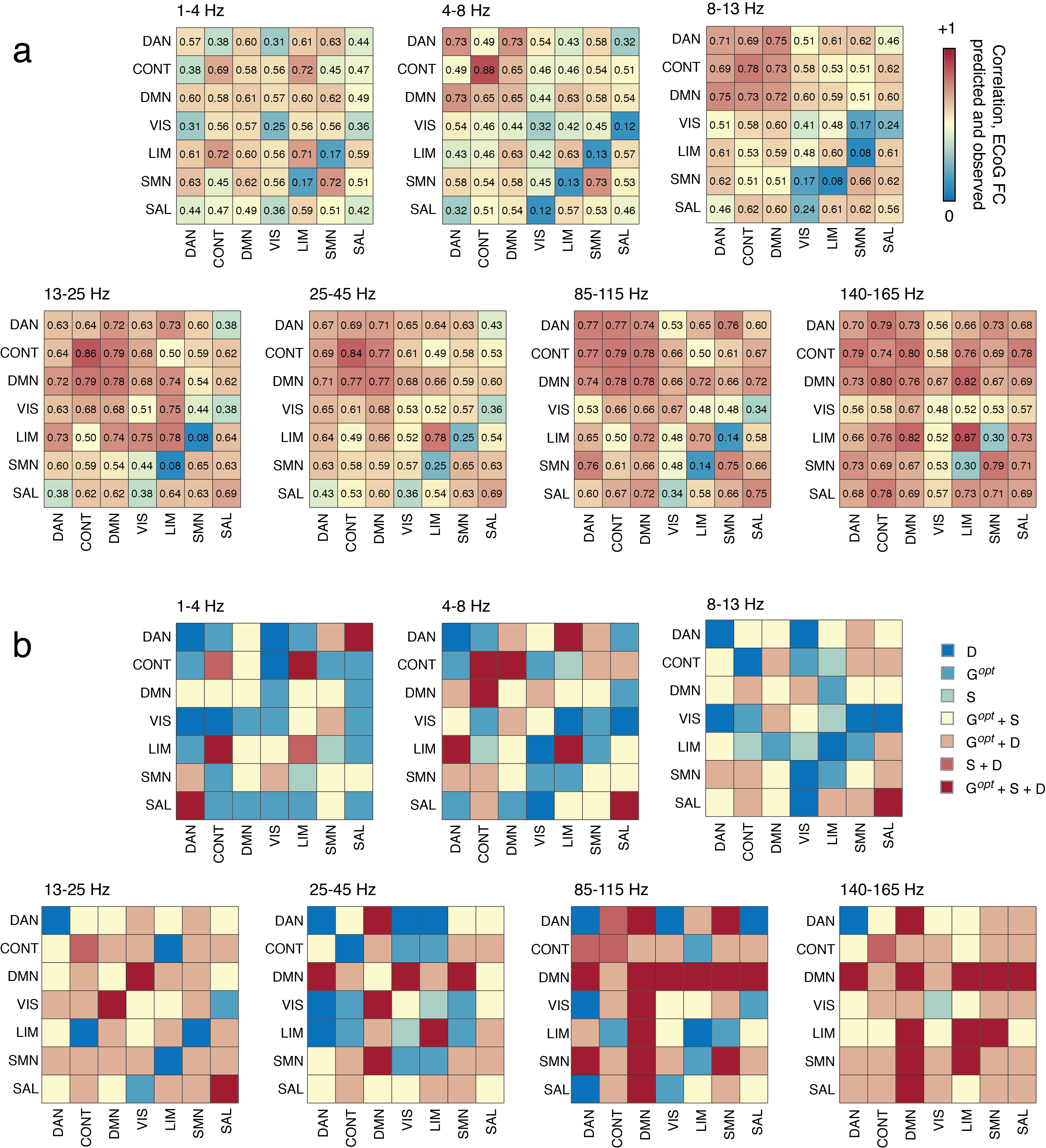}}
	\caption{\textbf{System-specific multi-linear models.} (\emph{a}) Model performance for connections within and between every pair of functional systems and across frequency bands. (\emph{b}) Optimal models, as identified using Akaike information criterion, for every pair of systems and across frequency bands.} \label{MLM_systems}
\end{center}
\end{figure*}

We repeated this procedure for each frequency band. Similar to the effect we observed in the main text, where model performance increased more or less monotonically with frequency, we found that overall model performance within specific blocks produced a similar frequency-specifc pattern (Fig.~\ref{MLM_systems}a). However, there was considerable heterogeneity across blocks and within a given frequency band, both in terms of model performance and the optimal model (as identified by AIC). For example, in the 1-4 Hz range, the correlation of observed and predicted ECoG FC within the limbic system was $r = 0.71$ ($p \approx 10^{-8}$). The optimal model, in that case, was one that included only interregional distance and search information (Fig.~\ref{MLM_systems}b). In contrast, the correlation of observed and predicted ECoG FC within the visual network was $r = 0.25$ ($p = 10^{-1.61}$) and based on a model that included optimized gene co-expression, alone. The variability in model performance suggests that some systems (those where ECoG FC is not well-predicted) may require additional predictors beyond those considered here. For example, brain regions can influence one another vascularly, possibly \emph{via} coordinated vasodilation \cite{warren2017surgically}.

\begin{figure*}[t]
\begin{center}
	\centerline{\includegraphics[width=0.75\textwidth]{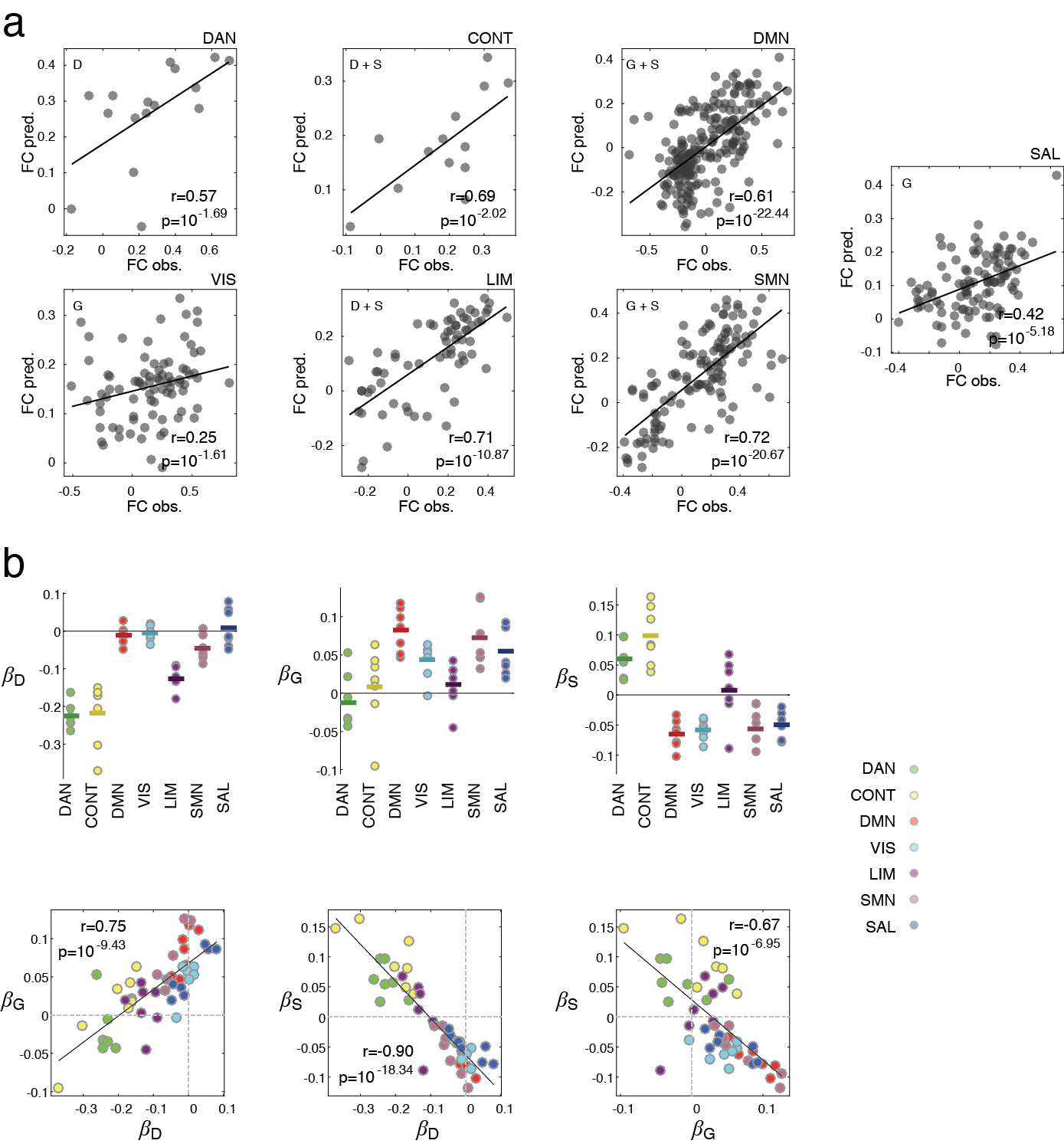}}
	\caption{\textbf{System-level model summary.} (\emph{a}) Scatterplots of models fit to within-system connections for the slowest (1-4 Hz) frequency band. (\emph{b}) Model regression coefficients aggregated across frequency bands reveals system heterogeneity. (\emph{c}) Model parameters, especially the Euclidean distance and search information regression coefficients ($\beta_D$ and $\beta_S$) appear to trade off with one another.} \label{MLM_systems_pt2}
\end{center}
\end{figure*}

While the MLM was fit to both the within- and between-system blocks of connections, the within-system blocks were of particular interest. Connections that fall among regions comprising the same system are thought to promote coordination and functional stability within the system \cite{sporns2016modular}. Accordingly, we focus on the within-system blocks. As before, model performance varied across systems (Fig.~\ref{MLM_systems_pt2}a). Beyond performance, however, we can also examine the model parameters themselves, $\beta_D$, $\beta_G$, and $\beta_S$. While we can analyze these parameters a number of different ways, we looked for parameters that maintained consistent sign across frequency bands, suggesting that that predictor contributes to the ECoG FC prediction in a consistent way. To ensure that each parameter existed for each system and for each frequency band, we focused on the full model that included all three predictors.

We observed that parameter values and their signs varied across systems, including some systems for which the parameters maintained consistent sign across all frequency bands, indicating that the corresponding predictors also play consistent roles and suggesting system-specific organizational principles (Fig.~\ref{MLM_systems_pt2}b). For example, ECoG FC within the dorsal attention and control networks are anti-correlated with distance and correlated with search information across all frequency bands (Fig.~\ref{MLM_systems_pt2}). This approach also makes it possible to identify systems that depend on gene expression profiles in a consistent way. For example, $\beta_G$ for visual, somatomotor, and salience networks maintain the same sign across frequency bands.

\begin{figure*}[t]
\begin{center}
	\centerline{\includegraphics[width=0.75\textwidth]{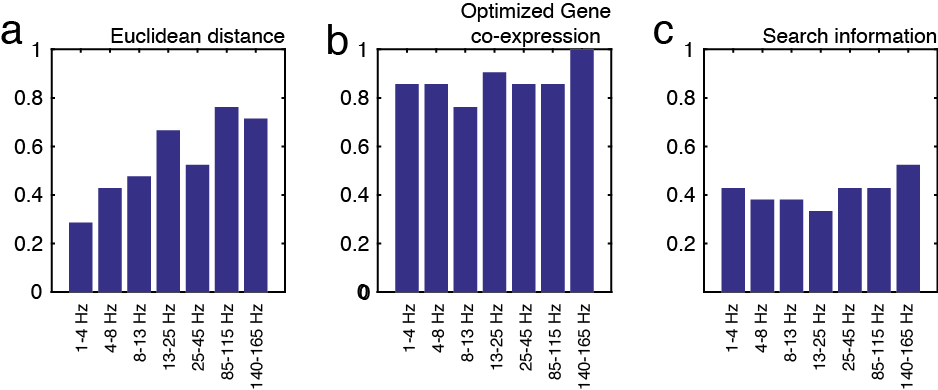}}
	\caption{\textbf{Fractional inclusion of specific predictors.} Fraction of system-level models that included Euclidean distance (\emph{a}), correlated gene expression (\emph{b}), and search information (\emph{c}) as predictors.} \label{modelFraction}
\end{center}
\end{figure*}

While the optimal model varied depending upon the system-by-system block to which the MLM was fit, many of the models overlapped in terms of the predictors they included. For example, the distance + genetics + search information model shares two predictors with the distance + search information model. We calculated the fraction of models that included each of the three predictors and repeated this procedure for all seven frequency bands. In the main text, we reported that the performance of the single-predictor distance model increased as a function of frequency, likely because long-distance ECoG FC is reduced in the faster bands. Mirroring this effect, we observed that the fraction of all models that include Euclidean distance as a predictor increased with frequency (Fig.~\ref{modelFraction}a). This trend was not clearly evident in either search information, which was included in approximately 40\% of models across frequency bands, nor in optimized gene co-expression, which was featured in approximately 80\% of models (Fig.~\ref{modelFraction}b,c).

While we see clear system-level differences in model performance, we resist over-interpreting these results. Part of the reason for not interpreting these results is that there is a disparity in the number of observed connections within \emph{versus} between systems. For example, while we have estimates for $\approx88\%$ (106/120) of the connections within the visual system, we have estimates for only $\approx47\%$ (17/36) of the connections within the control system. While it is possible that the missing connections are predicted with precisely the same weighted combinations of search information, correlated gene expression, and Euclidean distance as the observed connections, it is also possible that they are best-predicted by some different weighting of those factors. With greater numbers of subjects (and presumably increased cortical coverage) the number of missing connections decreases. Future work, therefore, will be directed at investigating system-level differences in MLMs.

\begin{table*}
	\begin{tabular}{|C{2cm}|C{1.5cm}|C{1.5cm}|C{1.5cm}|C{1.5cm}|C{1.5cm}|C{1.5cm}|C{1.75cm}|}
		\hline
		\textbf{Model} & $D$ & $G^{\text{R}}$ & $S$ & $G^{\text{R}},S$ & $G^{\text{R}},D$ & $S,D$ & $G^{\text{R}},D,S$ \\
		\hline
		1-4 Hz & $0.345$ & $0.186$ & $0.343$ & $0.372$ & $0.358$ & $0.363$ & $\textbf{0.381}$ \\
		\hline
		4-8 Hz & $0.401$ & $0.189$ & $0.390$ & $0.416$ & $0.409$ & $0.418$ & $\textbf{0.430}$ \\
		\hline
		8-13 Hz & $0.477$ & $0.230$ & $0.446$ & $0.477$ & $0.487$ & $0.489$ & $\textbf{0.502}$ \\
		\hline
		13-25 Hz & $0.547$ & $0.284$ & $0.487$ & $0.533$ & $0.562$ & $0.553$ & $\textbf{0.571}$ \\
		\hline
		25-45 Hz & $0.540$ & $0.291$ & $0.470$ & $0.526$ & $0.558$ & $0.545$ & $\textbf{0.566}$ \\
		\hline
		85-115 Hz & $0.604$ & $0.238$ & $0.553$ & $0.581$ & $0.606$ & $0.618$ & $\textbf{0.624}$ \\
		\hline
		140-165 Hz & $0.654$ & $0.258$ & $0.589$ & $0.621$ & $0.657$ & $0.667$ & $\textbf{0.674}$ \\
		\hline
	\end{tabular}
	\caption{\textbf{Model output using the ``Richiardi'' subset of genes to construct the correlation matrix of brain regions' gene expression profiles.} Each column represents one of seven multi-linear models. The first row indicates which measures were used as predictors: $D$, $G^{\text{R}}$, and $S$ represent Euclidean distance, gene co-expression from the Richiardi subset of genes, and search information. The next seven rows show the Pearson correlation magnitude of the predicted and empirical ECoG FC. The optimal models as determined by AIC are shown in boldface type.} \label{tables1}
\end{table*}

\clearpage

\begin{table*}
	\begin{tabular}{|C{2cm}|C{1.5cm}|C{1.5cm}|C{1.5cm}|C{1.5cm}|C{1.5cm}|C{1.5cm}|C{1.5cm}|C{1.75cm}|}
		\hline
		\textbf{Model} & $D$ & $G$ & $G^{\text{opt.}}$ & $S$ & $G^{\text{opt.}},S$ & $G^{\text{opt.}},D$ & $S,D$ & $G^{\text{opt.}},D,S$ \\
		\hline
		1-4 Hz & $0.345$ & $0.186$ & $0.523$ & $0.326$ & $0.537$ & $0.538$ & $0.363$ & $\textbf{0.540}$ \\
		\hline
		4-8 Hz & $0.401$ & $0.189$ & $0.512$ & $0.372$ & $0.538$ & $0.537$ & $0.419$ & $\textbf{0.543}$ \\
		\hline
		8-13 Hz & $0.477$ & $0.230$ & $0.563$ & $0.428$ & $0.592$ & $0.593$ & $0.491$ & $\textbf{0.599}$ \\
		\hline
		13-25 Hz & $0.547$ & $0.284$ & $0.578$ & $0.468$ & $0.622$ & $0.636$ & $0.556$ & $\textbf{0.641}$ \\
		\hline
		25-45 Hz & $0.540$ & $0.291$ & $0.557$ & $0.445$ & $0.604$ & $0.620$ & $0.546$ & $\textbf{0.624}$ \\
		\hline
		85-115 Hz & $0.604$ & $0.238$ & $0.546$ & $0.490$ & $0.614$ & $0.648$ & $0.610$ & $\textbf{0.653}$ \\
		\hline
		140-165 Hz & $0.654$ & $0.258$ & $0.578$ & $0.545$ & $0.659$ & $0.689$ & $0.666$ & $\textbf{0.699}$ \\
		\hline
	\end{tabular}
	\caption{\textbf{Model output using novel diffusion imaging dataset to construct search information matrix.} Each column represents one of eight multi-linear models. The first row indicates which measures were used as predictors: $D$, $G$, $G^{\text{opt}}$, and $S$ represent Euclidean distance, gene co-expression, optimized gene co-expression, and search information. The next seven rows show the Pearson correlation magnitude of the predicted and empirical ECoG FC. The optimal models as determined by AIC are shown in boldface type.} \label{tables3}
\end{table*}

\begin{table*}
	\begin{tabular}{|C{2cm}|C{1.5cm}|C{1.5cm}|C{1.5cm}|C{1.5cm}|C{1.5cm}|C{1.5cm}|C{1.5cm}|C{1.75cm}|}
		\hline
		\textbf{Model} & $D$ & $G$ & $G^{\text{opt.}}$ & $S$ & $G^{\text{opt.}},S$ & $G^{\text{opt.}},D$ & $S,D$ & $G^{\text{opt.}},D,S$ \\
		\hline
		1-4 Hz & $0.528$ & $0.099$ & $0.616$ & $0.498$ & $0.657$ & $0.660$ & $0.540$ & $\mathbf{0.664}$ \\
		\hline
		4-8 Hz & $0.552$ & $0.115$ & $0.596$ & $0.529$ & $0.651$ & $0.649$ & $0.567$ & $\mathbf{0.657}$ \\
		\hline
		8-13 Hz & $0.571$ & $0.176$ & $0.619$ & $0.520$ & $0.662$ & $0.667$ & $0.577$ & $\mathbf{0.671}$ \\
		\hline
		13-25 Hz & $0.605$ & $0.167$ & $0.633$ & $0.542$ & $0.687$ & $0.697$ & $0.610$ & $\mathbf{0.701}$ \\
		\hline
		25-45 Hz & $0.595$ & $0.164$ & $0.616$ & $0.523$ & $0.672$ & $0.680$ & $0.597$ & $\mathbf{0.684}$ \\
		\hline
		85-115 Hz & $0.654$ & $0.156$ & $0.591$ & $0.597$ & $0.681$ & $0.700$ & $0.662$ & $\mathbf{0.705}$ \\
		\hline
		140-165 Hz & $0.736$ & $0.174$ & $0.630$ & $0.661$ & $0.737$ & $0.766$ & $0.742$ & $\mathbf{0.771}$ \\
		\hline
	\end{tabular}
	\caption{\textbf{Model output as a result of fitting models to frequency-specific connection weights estimated for the same set of observed connections.} Each column represents one of eight multi-linear models. The first row indicates which measures were used as predictors: $D$, $G$, $G^{\text{opt}}$, and $S$ represent Euclidean distance, gene co-expression, optimized gene co-expression, and search information. The next seven rows show the Pearson correlation magnitude of the predicted and empirical ECoG FC. The optimal models as determined by AIC are shown in boldface type.} \label{tables4}
\end{table*}

\begin{table*}
	\begin{tabular}{|C{2cm}|C{8cm}|C{2cm}|}
		\hline
		\textbf{GO Term} & \textbf{Description} & \textbf{p-value} \\
		\hline
		GO:0005248 & voltage-gated sodium channel activity & $4.79 \times 10^{-4}$ \\
		\hline
		GO:1905030 & voltage-gated ion channel activity involved in regulation of postsynaptic membrane potential & $4.79 \times 10^{-4}$ \\
		\hline
	\end{tabular}
	\caption{\textbf{List of gene ontology terms, descriptions, and $p$-values associated with molecular function for optimized list of genes, 1-4 Hz.}} \label{freq1function}
\end{table*}

\begin{table*}
	\begin{tabular}{|C{2cm}|C{8cm}|C{2cm}|}
		\hline
		\textbf{GO Term} & \textbf{Description} & \textbf{p-value} \\
		\hline
		GO:0072111 & cell proliferation involved in kidney development & $1.34 \times 10^{-5}$ \\
		\hline
		GO:0006814 & sodium ion transport & $2.97 \times 10^{-5}$ \\
		\hline
		GO:0045880 & positive regulation of smoothened signaling pathway & $5.52 \times 10^{-5}$ \\
		\hline
		GO:0051239 & regulation of multicellular organismal process & $6.94 \times 10^{-5}$ \\
		\hline
		GO:0061138 & morphogenesis of a branching epithelium & $9.49 \times 10^{-5}$ \\
		\hline
		GO:0048754 & branching morphogenesis of an epithelial tube & $1.36 \times 10^{-4}$ \\
		\hline
		GO:0045650 & negative regulation of macrophage differentiation & $1.43 \times 10^{-4}$ \\
		\hline
		GO:0001763 & morphogenesis of a branching structure & $1.6 \times 10^{-4}$ \\
		\hline
		GO:0035239 & tube morphogenesis & $1.7 \times 10^{-4}$ \\
		\hline
		GO:0045649 & regulation of macrophage differentiation & $2.21 \times 10^{-4}$ \\
		\hline
		GO:0061820 & telomeric D-loop disassembly & $2.26 \times 10^{-4}$ \\
		\hline
		GO:0061004 & pattern specification involved in kidney development & $2.26 \times 10^{-4}$ \\
		\hline
		GO:0015672 & monovalent inorganic cation transport & $2.33 \times 10^{-4}$ \\
		\hline
		GO:0097306 & cellular response to alcohol & $2.38 \times 10^{-4}$ \\
		\hline
		GO:0061227 & pattern specification involved in mesonephros development & $2.65 \times 10^{-4}$ \\
		\hline
		GO:0072098 & anterior/posterior pattern specification involved in kidney development & $2.65 \times 10^{-4}$ \\
		\hline
		GO:0050976 & detection of mechanical stimulus involved in sensory perception of touch & $2.65 \times 10^{-4}$ \\
		\hline
		GO:0051049 & regulation of transport & $3.29 \times 10^{-4}$ \\
		\hline
		GO:0070723 & response to cholesterol & $3.32 \times 10^{-4}$ \\
		\hline
		GO:1902106 & negative regulation of leukocyte differentiation & $4.41 \times 10^{-4}$ \\
		\hline
		GO:0002062 & chondrocyte differentiation & $4.58 \times 10^{-4}$ \\
		\hline
		GO:0044707 & single-multicellular organism process & $4.61 \times 10^{-4}$ \\
		\hline
		GO:0032352 & positive regulation of hormone metabolic process & $4.72 \times 10^{-4}$ \\
		\hline
		GO:0090657 & telomeric loop disassembly & $4.72 \times 10^{-4}$ \\
		\hline
		GO:0032350 & regulation of hormone metabolic process & $5.68 \times 10^{-4}$ \\
		\hline
		GO:0036314 & response to sterol & $5.68 \times 10^{-4}$ \\
		\hline
		GO:0086010 & membrane depolarization during action potential & $5.77 \times 10^{-4}$ \\
		\hline
		GO:0001823 & mesonephros development & $6.42 \times 10^{-4}$ \\
		\hline
		GO:0072170 & metanephric tubule development & $6.42 \times 10^{-4}$ \\
		\hline
		GO:0072234 & metanephric nephron tubule development & $6.42 \times 10^{-4}$ \\
		\hline
		GO:0072243 & metanephric nephron epithelium development & $6.42 \times 10^{-4}$ \\
		\hline
		GO:0035725 & sodium ion transmembrane transport & $7.16 \times 10^{-4}$ \\
		\hline
		GO:0060537 & muscle tissue development & $7.16 \times 10^{-4}$ \\
		\hline
		GO:0046883 & regulation of hormone secretion & $7.17 \times 10^{-4}$ \\
		\hline
		GO:0061209 & cell proliferation involved in mesonephros development & $7.86 \times 10^{-4}$ \\
		\hline
		GO:0072138 & mesenchymal cell proliferation involved in ureteric bud development & $7.86 \times 10^{-4}$ \\
		\hline
		GO:0021773 & striatal medium spiny neuron differentiation & $7.86 \times 10^{-4}$ \\
		\hline
		GO:0051795 & positive regulation of timing of catagen & $7.86 \times 10^{-4}$ \\
		\hline
		GO:0044057 & regulation of system process & $7.97 \times 10^{-4}$ \\
		\hline
		GO:0051240 & positive regulation of multicellular organismal process & $8.05 \times 10^{-4}$ \\
		\hline
		GO:0001655 & urogenital system development & $8.45 \times 10^{-4}$ \\
		\hline7
		GO:0072207 & metanephric epithelium development & $8.45 \times 10^{-4}$ \\
		\hline
	\end{tabular}
	\caption{\textbf{List of gene ontology terms, descriptions, and $p$-values associated with biological processes for optimized list of genes, 1-4 Hz.}} \label{freq1process}
\end{table*}

\begin{table*}
	\begin{tabular}{|C{2cm}|C{8cm}|C{2cm}|}
		\hline
		\textbf{GO Term} & \textbf{Description} & \textbf{p-value} \\
		\hline
		GO:0005261 & cation channel activity & $6.55 \times 10^{-6}$ \\
		\hline
		GO:0005248 & voltage-gated sodium channel activity & $5.78 \times 10^{-5}$ \\
		\hline
		GO:1905030 & voltage-gated ion channel activity involved in regulation of postsynaptic membrane potential & $5.78 \times 10^{-5}$ \\
		\hline
		GO:0022832 & voltage-gated channel activity & $6.81 \times 10^{-5}$ \\
		\hline
		GO:0005244 & voltage-gated ion channel activity & $6.81 \times 10^{-5}$ \\
		\hline
		GO:0005216 & ion channel activity & $1.25 \times 10^{-4}$ \\
		\hline
		GO:0022836 & gated channel activity & $1.38 \times 10^{-4}$ \\
		\hline
		GO:0030547 & receptor inhibitor activity & $1.42 \times 10^{-4}$ \\
		\hline
		GO:0022838 & substrate-specific channel activity & $1.72 \times 10^{-4}$ \\
		\hline
		GO:0015267 & channel activity & $3.94 \times 10^{-4}$ \\
		\hline
		GO:0022803 & passive transmembrane transporter activity & $4.05 \times 10^{-4}$ \\
		\hline
		GO:0008092 & cytoskeletal protein binding & $4.9 \times 10^{-4}$ \\
		\hline
		GO:0005267 & potassium channel activity & $5.58 \times 10^{-4}$ \\
		\hline
		GO:0046873 & metal ion transmembrane transporter activity & $6.07 \times 10^{-4}$ \\
		\hline
		GO:0015079 & potassium ion transmembrane transporter activity & $7.81 \times 10^{-4}$ \\
		\hline
	\end{tabular}
	\caption{\textbf{List of gene ontology terms, descriptions, and $p$-values associated with molecular function for optimized list of genes, 4-8 Hz.}} \label{freq2function}
\end{table*}

\begin{table*}
	\begin{tabular}{|C{2cm}|C{8cm}|C{2cm}|}
		\hline
		\textbf{GO Term} & \textbf{Description} & \textbf{p-value} \\
		\hline
		GO:0044699 & single-organism process & $5.69 \times 10^{-6}$ \\
		\hline
		GO:0015672 & monovalent inorganic cation transport & $1.58 \times 10^{-5}$ \\
		\hline
		GO:0051899 & membrane depolarization & $2.24 \times 10^{-5}$ \\
		\hline
		GO:0030001 & metal ion transport & $2.46 \times 10^{-5}$ \\
		\hline
		GO:0006812 & cation transport & $5 \times 10^{-5}$ \\
		\hline
		GO:0044707 & single-multicellular organism process & $7.49 \times 10^{-5}$ \\
		\hline
		GO:0001508 & action potential & $8.04 \times 10^{-5}$ \\
		\hline
		GO:0045880 & positive regulation of smoothened signaling pathway & $1.08 \times 10^{-4}$ \\
		\hline
		GO:0086010 & membrane depolarization during action potential & $1.24 \times 10^{-4}$ \\
		\hline
		GO:0019228 & neuronal action potential & $1.86 \times 10^{-4}$ \\
		\hline
		GO:0043269 & regulation of ion transport & $2.31 \times 10^{-4}$ \\
		\hline
		GO:0035725 & sodium ion transmembrane transport & $3.12 \times 10^{-4}$ \\
		\hline
		GO:2000698 & positive regulation of epithelial cell differentiation involved in kidney development & $3.43 \times 10^{-4}$ \\
		\hline
		GO:0072139 & glomerular parietal epithelial cell differentiation & $3.52 \times 10^{-4}$ \\
		\hline
		GO:0050976 & detection of mechanical stimulus involved in sensory perception of touch & $3.52 \times 10^{-4}$ \\
		\hline
		GO:0048856 & anatomical structure development & $4.17 \times 10^{-4}$ \\
		\hline
		GO:0035994 & response to muscle stretch & $4.67 \times 10^{-4}$ \\
		\hline
		GO:0006814 & sodium ion transport & $4.67 \times 10^{-4}$ \\
		\hline
		GO:0044767 & single-organism developmental process & $6.07 \times 10^{-4}$ \\
		\hline
		GO:0060159 & regulation of dopamine receptor signaling pathway & $7.14 \times 10^{-4}$ \\
		\hline
		GO:0042391 & regulation of membrane potential & $7.74 \times 10^{-4}$ \\
		\hline
		GO:0000226 & microtubule cytoskeleton organization & $7.9 \times 10^{-4}$ \\
		\hline
		GO:0051049 & regulation of transport & $8.5 \times 10^{-4}$ \\
		\hline
		GO:0098662 & inorganic cation transmembrane transport & $8.54 \times 10^{-4}$ \\
		\hline
		GO:0044708 & single-organism behavior & $8.95 \times 10^{-4}$ \\
		\hline
		GO:0090257 & regulation of muscle system process & $9.17 \times 10^{-4}$ \\
		\hline
		GO:0006813 & potassium ion transport & $9.48 \times 10^{-4}$ \\
		\hline
		GO:0007010 & cytoskeleton organization & $9.66 \times 10^{-4}$ \\
		\hline
		GO:0009612 & response to mechanical stimulus & $9.99 \times 10^{-4}$ \\
		\hline
	\end{tabular}
	\caption{\textbf{List of gene ontology terms, descriptions, and $p$-values associated with biological processes for optimized list of genes, 4-8 Hz.}} \label{freq2process}
\end{table*}

\begin{table*}
	\begin{tabular}{|C{2cm}|C{8cm}|C{2cm}|}
		\hline
		\textbf{GO Term} & \textbf{Description} & \textbf{p-value} \\
		\hline
		GO:0034703 & cation channel complex & $2.49 \times 10^{-5}$ \\
		\hline
		GO:0001518 & voltage-gated sodium channel complex & $1.05 \times 10^{-4}$ \\
		\hline
		GO:0034702 & ion channel complex & $2.6 \times 10^{-4}$ \\
		\hline
		GO:1902495 & transmembrane transporter complex & $3.11 \times 10^{-4}$ \\
		\hline
		GO:0034705 & potassium channel complex & $3.12 \times 10^{-4}$ \\
		\hline
		GO:1990351 & transporter complex & $3.97 \times 10^{-4}$ \\
		\hline
		GO:0034706 & sodium channel complex & $4.67 \times 10^{-4}$ \\
		\hline
	\end{tabular}
	\caption{\textbf{List of gene ontology terms, descriptions, and $p$-values associated with cellular components for optimized list of genes, 4-8 Hz.}} \label{freq2component}
\end{table*}
	
\clearpage
\bibliography{biblio.bib, references.bib}

\end{document}